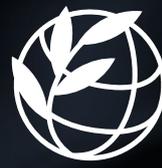

# Broadening the GIFCT Hash-Sharing Database Taxonomy: An Assessment and Recommended Next Steps

July 2021

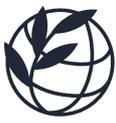

## Acknowledgements

The Global Internet Forum to Counter Terrorism would like to thank all of our stakeholders and partners for their interest in and commitment to this body of work. We are especially grateful to our GIFCT member companies, members of our Independent Advisory Committee (IAC), and civil society representatives for their feedback throughout this process, as well as to the GIFCT interns—Maggie Frankel, Aaron Tielemans, and Ye Bin Won—who provided significant support in the production of this series. GIFCT is also enormously appreciative of the time, expertise, and innovative thinking that all of the authors of the subsequent papers contributed to this project.

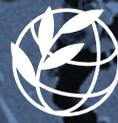

# Dynamic Matrix of Extremisms and Terrorism (DMET):

A Continuum Approach Towards Identifying Different Degrees of Extremisms

By Marten Risius, Kevin M. Blasiak, Susilo Wibisono, Rita Jabri-Markwell, and Winnifred Louis

# Dynamic Matrix of Extremisms and Terrorism (DMET)

## A Continuum Approach Towards Identifying Different Degrees of Extremisms

By Marten Risius, Kevin M. Blasiak, Susilo Wibisono, Rita Jabri-Markwell and Winnifred Louis

## Abstract
We propose to extend the current binary understanding of terrorism (versus non-terrorism) with a Dynamic Matrix of Extremisms and Terrorism (DMET). DMET considers the whole ecosystem of content and actors that can contribute to a continuum of extremism (e.g., right-wing, left-wing, religious, separatist, single-issue). It organizes levels of extremisms by varying degrees of ideological engagement and the presence of violence identified (e.g., partisan, fringe, violent extremism, terrorism) based on cognitive and behavioral cues and group dynamics. DMET is globally applicable due to its comprehensive conceptualization of the levels of extremisms. It is also dynamic, enabling iterative mapping with the region- and time-specific classifications of extremist actors. Once global actors recognize DMET types and their distinct characteristics, they can comprehensively analyze the profiles of extremist actors (e.g., individuals, groups, movements), track these respective actors and their activities (e.g., social media content) over time, and launch targeted counter activities (e.g. de-platforming, content moderation, or redirects to targeted CVE narratives).

## Key recommendations
1. Understand extremism as a dimensional concept with terrorism as a deviant pole from the regional norm.
2. Transparently map cues of extremism to accountably define categories of extremisms, creating an opportunity for dialogue between scholars and industry, and increasing trust with civil society and broader public audiences.
3. Provide the opportunity for organizations to upweigh or downweigh cues of extremism based on their local norms or national or international legal requirements, while changing the classification in an explainable and transparent way.
4. Iteratively update the manifestations of cues that characterize extremisms to account for changing profiles of extremism regionally or temporally.
5. Recognize all forms of violence used by violent extremists, especially serial or systematic dehumanization of an out-group as an attribute and indicator of violent extremism.
6. Enable platform providers to transparently decide and explain decisions to exempt extremist actors or content from DMET.
7. Create a more nuanced understanding of the degrees of extremism to reduce the probability of misclassifications (i.e., of non-terrorists as terrorists, or failing to identify terrorists as such) and allow more fine-grained analysis of actor changes over time.

## Introduction

Moderation of terrorists online is commonly achieved by referring to lists of known terrorist individuals and groups from academia, civil society, and governments.[51] These lists are very helpful for the differentiation between terrorist and non-terrorist content. Currently, the GIFCT hash-sharing database is an important tool for the list-based moderation of terrorist content online.

There are, however, certain issues that accompany these list-based approaches that remain to be solved. As summarized by the recent report from the Royal United Services Institute, no single type of list simultaneously can fulfill all of the following three criteria while still being economically feasible:[52]

1. **Ideological fairness:** equal opportunity for all entities to be classified as terrorist;
2. **Global applicability:** transcend regional borders; and
3. **Update frequency:** near real-time updates.

Furthermore, there is a gray area that poses various noteworthy delicate challenges. Terrorists often communicate non-violent content that falls outside the categories of **Imminent Credible Threat, Graphic Violence Against Defenseless People, Glorification of Terror Acts, Recruitment & Instruction.** These more subtle messages still help to further extremist causes when they are not directly captured by the hash database taxonomy. They use social media for fundraising purposes[53] or to affirm grievances, ideologies, and share humanitarian purposes (e.g., pictures of ISIS-affiliated doctors helping injured children) without calling for violence.[54] Furthermore, perceived overlap in content between violent extremists and (political) partisan actors, self-determination-based movements, or state-sponsored information campaigns raise legal and ethical questions in regard to appropriate treatment. Accordingly, the decision to add an actor and their content to a list of known terrorists in order to moderate their online presence sets a high bar that allows for considerable damage to occur beforehand, is associated with a strong stigmatization transforming the decision into a political issue and makes a revision of the decision following resocialization efforts unlikely. A transparent framework like the proposed Dynamic Matrix of Extremisms and Terrorism (DMET) is needed to respond to these challenges.

---

51 Chris Meserole and Daniel Byman, "Terrorist Definitions and Designations Lists: What Technology Companies Need to Know," Royal United Services Institute for Defence and Security Studies, (July 2019), https://rusi.org/explore-our-research/publications/special-resources/terrorist-definitions-and-designations-lists-what-technology-companies-need-to-know.
52 Meserole and Byman, "Terrorist Definitions and Designations Lists," 2.
53 Tom Keatinge and Florence Keen, "Social Media and (Counter) Terrorist Finance: A Fund-Raising and Disruption Tool," Studies in Conflict & Terrorism 42, no. 1-2 (2019); Tom Keatinge, Florence Keen, and Kayla Izenman, "Fundraising for Right-Wing Extremist Movements," The RUSI Journal 164, no. 2 (2019).
54 Roderick Graham, "Inter-Ideological Mingling: White Extremist Ideology Entering the Mainstream on Twitter," Sociological Spectrum 36, no. 1 (2016).

# Background

## The Dynamic Matrix of Extremisms and Terrorism (DMET)

We propose extending the binary understanding of terrorism (versus non-terrorism) with a Dynamic Matrix of Extremisms and Terrorism (DMET) to address the intersection of extremism types (and associated extremist content). DMET identifies varying degrees of ideological engagement and violence based on cognitive and behavioral cues as well as group dynamics. [55]

## DMET's Understanding of Extremism

DMET understands extremism on a continuum of varying degrees of ideological engagement. Any label of "extremism" assigned in reference to DMET needs to share the spirit of the following assumptions that accompany DMET's continuum-based understanding.

First, the associated types of extremisms are based on an understanding of online extremism as a deviation from something that is commonly considered (more) "ordinary," "mainstream," or "normal."[56] DMET declines an evaluative notion of the purposes and goals of the different forms of extremism.

Second, the levels of ideological engagement are derived from an understanding of radicalization as the "change in beliefs, feelings, and behaviors in directions that increasingly justify intergroup violence and demand sacrifice in defense of the group."[57] Consequently, we focus on cognitive, behavioral, and group dynamic cues to describe the different levels of ideological engagement. These general descriptions then need to be operationalized respective to the regional and temporal context.

Third, DMET emphasizes the plurality of extremisms to underscore our assumption that extremism is a concept of varying degrees and deviation from regionally dominant ideologies. We emphasize this to avoid stigmatization of minorities as "extremists" for proposing views that deviate from the regional majority (e.g., Radical Veganism). Higher levels of ideological engagement in different forms of extremisms are characterized by more uncommon cognitive, behavioral, and group dynamic cues.

---

55 Peter R. Neumann, "The Trouble with Radicalization," International Affairs 89, no. 4 (2013); Kris Christmann, "Preventing Religious Radicalisation and Violent Extremism: A Systematic Review of the Research Evidence," Youth Justice Board (2012); Susilo Wibisono, Winnifred R Louis, and Jolanda Jetten, "A Multidimensional Analysis of Religious Extremism," Frontiers in Psychology, 10 (2019).
56 Alex P. Schmid, "Violent and Non-Violent Extremism: Two Sides of the Same Coin," International Centre for Counter-Terrorism (ICCT) Research Paper (2014); Charlie Winter et al., "Online Extremism: Research Trends in Internet Activism, Radicalization, and Counter-Strategies," International Journal of Conflict and Violence (IJCV) 14, no. 2 (2020): 4; Ronald Wintrobe, Rational Extremism: The Political Economy of Radicalism, Cambridge, New York, Cambridge University Press, (2006).
57 Clark McCauley and Sophia Moskalenko, "Mechanisms of Political Radicalization: Pathways toward Terrorism," Terrorism and Political Violence 20, no. 3 (2008); Winter et al., "Online Extremism."

Fourth, we consider the matrix to be dynamic to acknowledge both that groups change and that the understanding of "normal" is variable across geography and time. For example, a group that opposes vaccination may be viewed as fringe or extremist (i.e., non-normative) in certain parts of the world at a particular time, but regarded as normal in other parts or at other times.[58] Hence, assessments of ideological engagement and the underlying forms of operationalizations are regionally delimited, time-specific, and require regular updates.

## Level Defining Cues

DMET distinguishes between four levels of ideological engagement: partisan, fringe, violent extremist, and terrorist. DMET's continuum approach adopts the idea of ordering degrees of violent extremism[59] to extend the simplified categories of terrorism and non-terrorism. In DMET's case, we start from a point at the normative or moderate baseline and move through to increasing degrees of alienation from the mainstream to ultimately active violent acts. Each level of ideological engagement is proposed to have a particularly prevalent configuration of cues to identify and classify a group or content (i.e., from partisanship to terrorism). These cues are cognitive, behavioral, and group dynamic. A discussion of strategic and technical implementation considerations can be found in sections 4.2 and 4.3.

### Cognitive Cues in DMET

Cognitive cues are signals indicating the thoughts and attitudes of individuals or groups.[60] At the individual level, cognitive cues might emerge in the form of thoughts or images. At the collective or group level, cognitive cues are shared beliefs or representations involved in recognizing and perceiving ourselves and other individuals or social categories. Many outcomes can flow from these socio-cognitive processes, such as prejudice and stereotypes.[61] For the purpose of classifying content as extremist or not, key aspects tracked by DMET cognitively would include beliefs about or representations of one's own groups (ingroups) and their actions, the targeted opponent groups (out-groups) and their actions, the nature of right or wrong, and the nature of the threats or value differences that define the relationship between the groups.

---

58 Ayodele Samuel Jegede, "What Led to the Nigerian Boycott of the Polio Vaccination Campaign?," PLoS Medicine 4, no. 3 (2007).
59 For an example, see Donald Holbrook, "Designing and Applying an 'Extremist Media Index,'" Perspectives on Terrorism 9, no. 5 (2015).
60 Bert N. Bakker, Yphtach Lelkes, and Ariel Malka, "Understanding Partisan Cue Receptivity: Tests of Predictions from the Bounded Rationality and Expressive Utility Perspectives," The Journal of Politics 82, no. 3 (2020).
61 M. Verkuyten and A. De Wolf, "The Development of in-Group Favoritism: Between Social Reality and Group Identity," Developmental Psychology 43, no. 4 (2007).

46

## Behavioral Cues in DMET

Behavioral cues refer to the observable actions by groups or individuals or representations of those actions. At an individual level, behavioral cues can be observed from (for example) facial expression, gesture, vocal expression, etc.[62] For the purposes of DMET, these cues might indirectly identify (for example) the emotional level (e.g., anger) that a person has in one situation, or may explicitly show harm-doing and calls to violence. Behavioral cues can be addressed to the self or others. At a group level, drawing on the literature on political contestation, collective action,[63] and intergroup violence. For the purposes of DMET, we are most interested in coding for content that involves a call to cooperate with in-group or prospective allies or to engage in concrete actions that derogate or harm another group.

## Group Dynamics in DMET

The process of radicalization or increasing extremism often draws on group dynamics by establishing norms about appropriate and deviant behaviors, with very little latitude in accepting differences.[64] People may be drawn to identify with causes or groups based on broad in-/out-group dynamics, as intergroup threats and conflicts of interest or values are contested using a range of tactics, from debate and satire to threats, dehumanization, and violence.[65] The dynamic influence of group identities and norms can provide ideological glue for (de)radicalization across the extremist spectrum.[66]

Group dynamics refers to a system of behaviors and psychological processes occurring within or between social groups.[67] Intragroup dynamics (i.e., how individuals in a group interact with one another) underlie social processes that give rise to a set of norms, roles, relations, and common goals characterizing a particular group.[68] Group dynamics can also involve the cooperation or competition of individuals within the groups to gain group recognition or act on behalf of the group. In addition, intergroup dynamics (i.e., how groups interact with each other) include collective perception, attitudes, and actions

---

[62] Alessandro Vinciarelli et al., "Social Signal Processing: State-of-the-Art and Future Perspectives of an Emerging Domain," in Proceedings of the 16th ACM international conference on Multimedia (Vancouver, British Columbia, Canada: Association for Computing Machinery, 2008).
[63] Defined as any action aimed to improve the group's status; see M. van Zomeren, T. Postmes, and R. Spears, "Toward an Integrative Social Identity Model of Collective Action: A Quantitative Research Synthesis of Three Socio-Psychological Perspectives," Psychological Bulletin 134, no. 4 (2008); S. C. Wright, D. M. Taylor, and F. M. Moghaddam, "Responding to Membership in a Disadvantaged Group - from Acceptance to Collective Protest," Journal of Personality and Social Psychology 58, no. 6 (1990).
[64] Wibisono, Louis, and Jetten, "A Multidimensional Analysis of Religious Extremism."
[65] C. Stott, P. Hutchison, and J. Drury, "'Hooligans' Abroad? Inter-Group Dynamics, Social Identity and Participation in Collective 'Disorder' at the 1998 World Cup Finals," British Journal of Social Psychology, 40 (2001).
[66] John M. Berger, "Deconstruction of Identity Concepts in Islamic State Propaganda: A Linkage-Based Approach to Counter-Terrorism Strategic Communications," The Hague, Netherlands: EUROPOL, (2017); Donald Holbrook, "Far Right and Islamist Extremist Discourses: Shifting Patterns of Enmity," Extreme Right Wing Political Violence and Terrorism (2013).
[67] M. A. Hogg and D. J. Terry, "Social Identity and Self-Categorization Processes in Organizational Contexts," Academy of Management Review 25, no. 1 (2000); J. Sidanius et al., "Ethnic Enclaves and the Dynamics of Social Identity on the College Campus: The Good, the Bad, and the Ugly," Journal of Personality and Social Psychology 87, no. 1 (2004).
[68] M. A. Hogg and S. A. Reid, "Social Identity, Self-Categorization, and the Communication of Group Norms," Communication Theory 16, no. 1 (2006).

toward other groups.[69]

Intra- and intergroup dynamics can produce and be shaped by specific behavioral and cognitive cues. However, the DMET coding here refers to specific attributes of how content is being disseminated relationally (e.g., conformity, polarization), and how sources are positioning themselves in relation to other groups (e.g., as leaders, warriors) and groups in relation to each other (e.g., as enemies, allies, dupes). Source attributes where available would be coded in Group Dynamics, both in terms of membership in particular groups and of position within particular networks (e.g., contact with a known violent actor).

## Types of Ideological Engagement

Crossed with these levels in DMET (see Figure 1), we consider five categories of actors/content according to their ideological arena: Right-Wing (e.g., concerning threats to the "white race" or "traditional values"), Left-Wing (e.g., concerning the need for a fair distribution of wealth), Religious (e.g., seeking to spread one's religion or purify it), Separatist (e.g., seeking territory for one's group), and Single-issue (e.g., advocating for one particular topic such as abortion or animal justice).[70] A group may be classified into more than one type of ideology, as it advocates for an issue by drawing narratively on other content (e.g., both right-wing ideology and religion). The purpose of the categories is to a) signal the inclusivity of DMET with all groups equally able to be considered as violent actors or terrorists; and b) build an understanding of how clusters of particular indicators or attributes emerge in different causes, resulting in profiles of domain-specific indicators feeding into context-specific categorization algorithms.

---

69 "Hooligans' Abroad?."
70 Allard R. Feddes et al., Psychological Perspectives of Radicalization (London: Routledge, 2020).

## The Continuum of Ideological Engagement

A core premise of DMET is its understanding of ideological engagement as a spectrum of varying degrees of severity (Figure 2) instead of the current binary dichotomy of (non) terrorism (Figure 1).

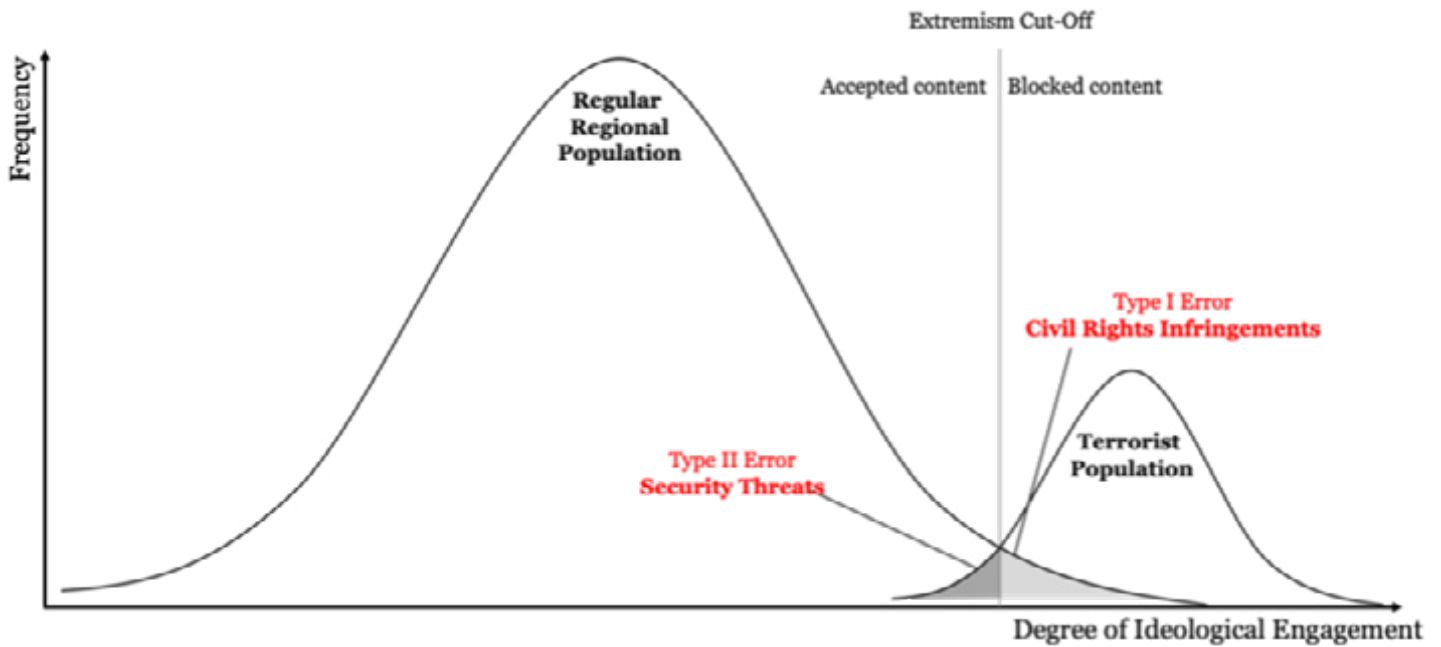

Figure 1. Conceptualization of Current Dichotomous Understanding of Ideological Engagement

This continuum perspective enables DMET to distinguish among different levels of ideological engagement (i.e., partisanship, fringe, violent extremism, terrorism) and the regular population norms that define the regionally accepted social standard. Thereby, the aim is to enable platform providers to make independent content or actor moderation decisions in a more nuanced fashion (including determining and disclosing cut-off values), with fewer misclassification errors between regular content and terrorist material. DMET also enables greater transparency regarding moderation decisions (e.g., by providing a means not only to classify organizations among the dimensions but also to develop and explain the weighing of attributes in the decision-making algorithms).

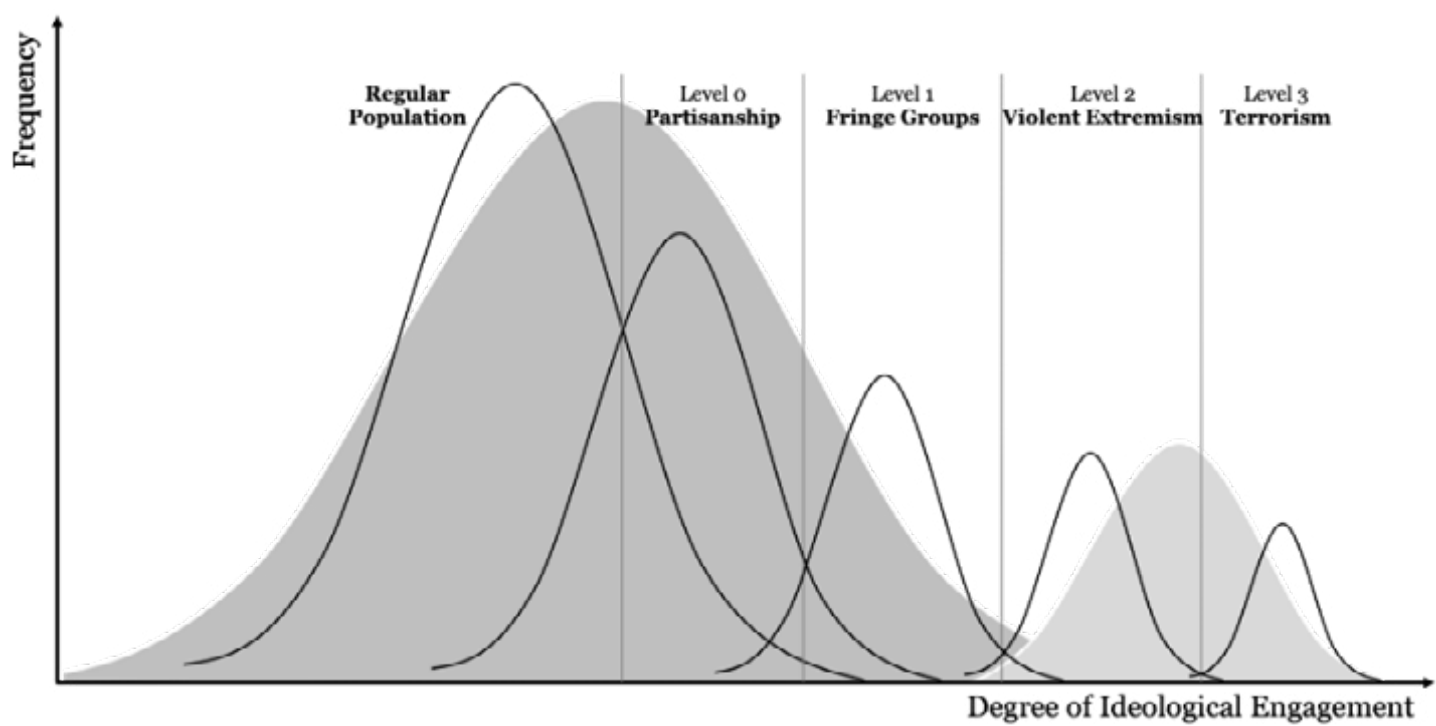

Figure 2. Conceptualization of DMET's Proposed Level's of Extremism Based on the Assumed Continuum of Ideological Engagement

It needs to be noted that the conceptualization of the different levels of ideological engagement as normally distributed sub-groups is a proposition that needs to be empirically tested following DMET's operationalization criteria outlined in the following. We also emphasize that this is meant as a conceptual illustration of the continuum of extremism and that the proportions of the more extreme sub-populations are likely smaller in reality.[71]

---

71 Dirk Oegema and Bert Klandermans, "Why Social Movement Sympathizers Don't Participate: Erosion and Nonconversion of Support," American Sociological Review 59, no. 5 (1994).

| Levels of Ideological Engagement | Level Defining Cues | | | Types of Ideological Engagement | | | | |
|---|---|---|---|---|---|---|---|---|
| | Cognitive | Behavioral | Group Dynamic | Right-Wing | Left-Wing | Religious | Separatist | Single-Issue |
| Terrorism | | | | | | | | |
| **Level 3** Terrorism | Sidestep inhibitory mechanisms, perceive target as 'the enemy,' legitimizing and valorizing death, wanting to intimidate broader population | Endorse, promote, or enact physical violence towards out-, in-group or infrastructure | Propagate values of active martyrdom, divide group labor to support violent acts | Ultra-right, far-right, alt-right, right-wing extremism, fascism, white supremacy | Ultra-left, far-left, left-wing extremism | Religiously motivated terrorism | Violent militant separatist organizations | Violent militant activism |
| Violent Extremism | | | | | | | | |
| **Level 2** Violent Extremism | Be intolerant towards others, represent cultural & structural violence through silencing and exclusion, perceive a reduced level of moral duties owed to the out-group | Serially or systematically dehumanize others, frequently express hate speech towards opponents, perform selective/individual acts of violence, actively separate targets from society, active discrimination | Compete for within-group recognition, show personal agency in the service of group domination, coalescing around out-group as a perceived or designated existential threat | Radical right, extreme conservatism | Radical left | Religiously motivated extremism | Secessionism, autonomism | Propaganda groups |
| Non-Violent Extremism | | | | | | | | |
| **Level 1** Fringe Group | Perceive/glorify the in-group as superior, indoctrinate dogmatic values, prejudice, and discrimination | Discredit or denigrate the out-group, seek isolation from the general public, express external blame for negative events, censor deviant views | Pursue and promote norms of purity, supremacy, domination, or revenge | Right-wing nationalist | Left-wing nationalist | Religious fundamentalism, cult | Seeking self-determination | Conspiracy theorists, fringe party advocating single-issue |
| Non-Extremism | | | | | | | | |
| **Level 0** Partisanship | Holding polarized and normative views, self-identifying with one group in opposition to another group | Expressing populist ideology, dog-whistling, satirizing other views, evangelizing others, campaigning peacefully | Holding political grievances, experiencing a sense of victimization or identity crises, or a need for significance | Right-wing populism | Left-wing populism, liberalism, socialism | Religious conservatism | Regional advocacy groups | Special-interest advocacy groups, lobbyism |

Please note: The table is based on a value-free understanding of extremism as something that is significantly deviant from the 'mainstream' or 'normal'; dynamic classifications of content, individuals or groups are dependent on the understanding of what is 'normal' in a particular region at a given point in time.

**Table 1. Dynamix Matrix of Extremisms and Terrorism (DMET)**

# Operationalization of DMET

In the following, we describe DMET's operationalization of the proposed levels of ideological engagement based on cognitive, behavioral, and group dynamic cues. We acknowledge the wealth of literature discussing comprehensive definitions of these concepts. For the sake of this briefing paper, we limit ourselves to deriving working definitions that explain DMET-based actor classifications.

We identify an indicative basket of indicator attributes for each level. However, part of our approach is that the association of any one attribute with a level or ideological cause (e.g., right-wing extremism) is dynamic and may change over time, so indicators wax or wane in their diagnostic value in historical periods or for particular contexts. Regional expert feedback to set the starting parameters and automated updating of the model over time will be important in sustaining DMET's accuracy.

## Level 0: Partisanship

Partisanship constitutes a non-extremist form of coordinated ideological engagement where individuals are committed to similar normative ideas and face conditions of conflict opposing others with whom they are at odds.[72] Partisans distinguish themselves from mainstream views through their normative ideology and offer a support network where collective actors are empowered to contest perceived grievances. The partisan commitment also serves as a source of identification and shapes the individuals' self-concept,[73] which leads to the continued endorsement of the mission that the group embodies and sustains the long-term pursuit of projects across a range of conditions and circumstances. Partisanship is not limited to in-group conformity but also is associated with perceived polarization away from a rival or opponent out-group.[74]

At a behavioral level, partisans commit to enacting a form of regulated adversarialism, which describes their commitment to persuade and evangelize others of their views tempered by self-set rules or ideals.[75] They may pursue different strategies, such as expressing populist ideologies as an opposition force or from a position of power.[76] Traditionally partisan actions take place through conventions, meetings, assemblies, and peaceful protests complemented by the online sphere via websites, blogs, and social media. Calls for action against the out-group are often indirect, using appeals that subtly invoke negative stereotypes about an opposing group (e.g., dog-whistling or racial priming theory) to harness the power of prejudice.[77] Partisans often target mainstream audiences, satirizing others by embedding ideological information into entertaining formats to engage others who are otherwise agnostic about a particular issue.[78]

---

72 Jonathan White and Lea Ypi, The Meaning of Partisanship, Oxford, Oxford University Press, (2016).
73 Emily A. West and Shanto Iyengar, "Partisanship as a Social Identity: Implications for Polarization," Political Behavior (2020).
74 Noam Lupu, "Party Polarization and Mass Partisanship: A Comparative Perspective," Political Behavior 37, no. 2 (2015).
75 White and Ypi, The Meaning of Partisanship.
76 S. Erdem Aytaç, Ali Çarkoğlu, and Ezgi Elçi, "Partisanship, Elite Messages, and Support for Populism in Power," European Political Science Review 13, no. 1 (2021).

In terms of group dynamics, partisans express and market feelings of injustice, grievances, or disaffection,[79] invoking personal and collective needs for significance and a desire to matter and be respected.[80] Narratives often identify an identity crisis threatening the group[81] and victimization at the hands of other out-groups.

## Level 1: Fringe Groups

According to DMET, fringe groups describe non-violent ideologies that are on the periphery of social movements or larger organizations, with more extreme views than those of the majority. Again, we acknowledge the considerable heterogeneity of ways to be a fringe actor and also the reality that in particular contexts, the toxic dynamics we ascribe below to "fringe" organizations may also apply to mainstream partisan groups. Based on this logic, we propose that DMET platforms create the opportunity to transparently and accountably "dial down" the diagnostic weighing of a particular dimension (e.g., out-group derogation) to avoid false positives when such rhetoric characterizes mainstream discourse.

With that caveat noted, DMET proposes that fringe groups are marked by cognitive cues such as beliefs of in-group superiority, out-group distinctiveness and inferiority, dogmatic values, learned prejudice, and discrimination.[82]

Behaviorally, we conceive that fringe groups discredit or denigrate the out-group, promote isolation from the general public, and promote narratives of external blame for negative outcomes such as conspiracy theories.[83]

---

77 Rachel Wetts and Robb Willer, "Who Is Called by the Dog Whistle? Experimental Evidence That Racial Resentment and Political Ideology Condition Responses to Racially Encoded Messages," Socius 5 (2019).
78 Silvia Knobloch-Westerwick and Simon M. Lavis, "Selecting Serious or Satirical, Supporting or Stirring News? Selective Exposure to Partisan Versus Mockery News Online Videos," Journal of Communication 67, no. 1 (2017).
79 Donald R. Kinder and D. Roderick Kiewiet, "Economic Discontent and Political Behavior: The Role of Personal Grievances and Collective Economic Judgments in Congressional Voting," American Journal of Political Science (1979); White and Ypi, The Meaning of Partisanship.
80 Arie W. Kruglanski et al., "The Psychology of Radicalization and Deradicalization: How Significance Quest Impacts Violent Extremism," Political Psychology 35 (2014).
81 John Sides, Michael Tesler, and Lynn Vavreck, Identity Crisis: The 2016 Presidential Campaign and the Battle for the Meaning of America (Princeton University Press, 2019).
82 Roy F. Baumeister, Evil: Inside Human Cruelty and Violence (WH Freeman/Times Books/Henry Holt & Co, 1996); Robert J. Sternberg, "A Duplex Theory of Hate: Development and Application to Terrorism, Massacres, and Genocide," Review of General Psychology 7, no. 3 (2003).
83 Marc W. Heerdink et al., "Emotions as Guardians of Group Norms: Expressions of Anger and Disgust Drive Inferences About Autonomy and Purity Violations," Cognition and Emotion 33, no. 3 (2019).

In turn, the group dynamics of fringe actors are marked by internal intolerance and censorship of deviant views, as well as readiness to pursue and promote norms of purity, supremacy, domination, or revenge.[84] Group members are socialized and indoctrinated into binary right–wrong classifications, sometimes in a highly systematic fashion in which newcomers move from the ideological periphery of their group to the inside through contracts of commitment and conversion, and in concert to withdrawing in isolation from other sources of identity such as family.[85]

## Level 2: Violent Extremism

Violent Extremists propagate a radical ideology supported by violent means that condone physical or mental harm to others. A key definitory factor that determines violent extremism is ideologically sanctioned violence such as dehumanization. Our concept is that groups often differ internally in the tactics advocated and contest the use of violence, and we seek to distinguish fringe groups in which isolated and/or peripheral members advocate for hate or violence from violent extremist groups where leaders and mainstream advocates do so, to terrorist groups where a formal division of labor to carry out attacks has been implemented.

On a cognitive level, violent extremists are intolerant towards others, representing cultural and structural violence through silencing and exclusion as just, inevitable, or appropriate, perceiving a reduced level of moral duties owed to the out-group. Extremists develop narratives legitimizing violence, often by framing the out-group as an enemy who is violent towards them.[86]

Behaviorally, violent extremists serially or systematically dehumanize others. Violent extremists refuse to tolerate or respect opinions or beliefs contrary to their own; they perceive a moral superiority and obligation to enforce their ideology.[87] This also frees extremists to act violently against the "other" without moral obligations and the burden of guilt that would typically be associated with violence.[88] Against that backdrop, these groups frequently express hate speech towards opponents to create psychological and structural violence through silencing and exclusion. Individual members of violent extremist groups may perform selective acts of physical violence as part of a group dynamic that valorizes these actions. Hate speech and glorification of violent acts would both be indicators of this level of engagement in DMET.[89]

---

84 Dominic Abrams et al., "Pro-Norm and Anti-Norm Deviance within and between Groups," Journal of Personality and Social Psychology 78, no. 5 (2000); Dominic Abrams et al., "Collective Deviance: Scaling up Subjective Group Dynamics to Superordinate Categories Reveals a Deviant Ingroup Protection Effect," Journal of Personality and Social Psychology (2021); Roger Giner-Sorolla, Bernhard Leidner, and Emanuele Castano, "Dehumanization, Demonization, and Morality Shifting: Paths to Moral Certainty in Extremist Violence," Extremism and the Psychology of Uncertainty (2012).
85  Andrew Coulson, "Education and Indoctrination in the Muslim World," Policy Analysis 29 (2004); Michael A. Hogg, Arie Kruglanski, and Kees Van den Bos, "Uncertainty and the Roots of Extremism," Journal of Social Issues 69, no. 3 (2013); F. M. Moghaddam, "The Staircase to Terrorism a Psychological Exploration;" American Psychologist 60, no. 2 (2005); John G. Horgan et al., "From Cubs to Lions: A Six Stage Model of Child Socialization into the Islamic State," Studies in Conflict & Terrorism 40, no. 7 (2017).
86 Douglas Pratt, "Religion and Terrorism: Christian Fundamentalism and Extremism," Terrorism and Political Violence 22, no. 3 (2010); David Webber and Arie W. Kruglanski, "The Social Psychological Makings of a Terrorist," Current Opinion in Psychology 19 (2018).
87 Hogg, Kruglanski, and Van den Bos, "Uncertainty and the Roots of Extremism."
88 Erving Goffman, Stigma: Notes on the Management of Spoiled Identity (Simon and Schuster, 2009); Albert Bandura, "Moral Disengagement in the Perpetration of Inhumanities," Personality and Social Psychology Review 3, no. 3 (1999).

At the group level, members of violent extremist groups compete for within-group recognition, seeking to show personal agency in the service of group domination, and coalescing around out-groups as perceived or designated existential threats.[90] The normative context of dehumanization establishes social preconditions within which violence by extremist instigators is likely to be perceived as justified. They authorize individuals to perform violence and shape bystanders' reactions to these events, while establishing the parameters for depersonalization and stigma or dehumanization and moral exclusion.[91] While these group dynamics might not be transparent at the content level, favorable responses valorizing particular in-group actors who are violent may provide a key set of indicators that would serve to identify the dynamics at play.[92]

## Level 3: Terrorism

Terrorism constitutes the most extreme form of ideologically driven engagement that uses violence even towards non-combatant targets to instill terror or to send a 'message'.[93] At the cognitive level, terrorists experience two key psychological processes involving a rigid, exclusive social categorization (e.g., of civilians as part of the out-group) and a greater psychological or moral distance by exaggerating differences between the in-group and the out-group.[94] The categorization of society at large as part of the out-group and as the enemy then serves as the justification for their struggle to intimidate or harm civilians.[95] Terrorists thereby sidestep "inhibitory mechanisms" that would normally limit the aggression of humans against one another. Instead, they show the greatest adherence to principles that move them to conform unconditionally to certain moral duties, which ultimately legitimize and valorize death.[96] Cognitive beliefs about the legitimacy of killing and the glory of risking sacrificial death are often indicators of the terrorist level in DMET.

---

89 Alexandra Olteanu et al., "The Effect of Extremist Violence on Hateful Speech Online" (paper presented at the Proceedings of the International AAAI Conference on Web and Social Media, 2018); Pratt, "Religion and Terrorism."
90 Gary A. Ackerman, Jun Zhuang, and Sitara Weerasuriya, "Cross-Milieu Terrorist Collaboration: Using Game Theory to Assess the Risk of a Novel Threat," Risk Analysis 37, no. 2 (2017); Randy Borum, "Radicalization into Violent Extremism I: A Review of Social Science Theories," Journal of Strategic Security 4, no. 4 (2011); David R. Mandel, "The Role of Instigators in Radicalization to Violent Extremism," Psychosocial, Organizational, and Cultural Aspects of Terrorism: Final Report to NATO HFM140/RTO. Brussels: NATO (2011).
91 Erving Goffman, Stigma: Notes on the Management of Spoiled Identity (Simon and Schuster, 2009); Albert Bandura, "Moral Disengagement in the Perpetration of Inhumanities," Personality and Social Psychology Review 3, no. 3 (1999).
92 Arie W. Kruglanski et al., "To the Fringe and Back: Violent Extremism and the Psychology of Deviance," American Psychologist 72, no. 3 (2017); Jeff Victoroff, Janice R. Adelman, and Miriam Matthews, "Psychological Factors Associated with Support for Suicide Bombing in the Muslim Diaspora," Political Psychology 33, no. 6 (2012); Webber and Kruglanski, "The Social Psychological Makings of a Terrorist."
93 Meserole and Byman, "Terrorist Definitions and Designations Lists."
94 Moghaddam, "The Staircase to Terrorism."
95 Marc Sageman, Understanding Terror Networks (University of Pennsylvania Press, 2011).

Behaviorally, terrorists also endorse, promote, and engage in violent and destructive actions. These are predominantly directed at civilians as well as non-human symbolic or infrastructure targets (e.g. works of art, places of worship).[97] Terrorists target different objectives depending on the specific sources of support available to them and the degree of out-group antagonism in their constituency.[98] Terrorists, however, also engage in violent actions against in-group members as (potential) defectors to sustain the long-term mission and group norms.[99] Concrete incitement to violence and physically violent acts provide behavioral indicators of the terrorist level in DMET.

In terms of group dynamics, terrorists have organized social structures that support violent actions on an ongoing basis. Terrorists often are taught to internalize the glorification of active martyrdom as a testimony of ideological commitment and faith.[100] We refer to active martyrdom (as opposed to passive martyrdom) as a characteristic of terrorism in the sense of a suicide attack where the act of self-destruction targets a perceived out-group enemy. Passive martyrdom (e.g., in politics or religion), where the actor is compelled to be ready to give one's life to defend the ideals and values of the group, is not considered a terrorist attribute.[101] However, while these beliefs are in theory shared by all group members, in practice, a formal division of labor to support violent acts exists (e.g., consisting of finances, military affairs, religious affairs, and public relations).[102]

In some cases, demographic divisions stream actors to different roles (e.g., younger men might be expected to serve as martyrs while older men direct actions and women serve support roles). In other cases, core groups of strategists and recruiters with ongoing roles might engage opportunistically, at a distance, with individuals of any age and gender recruited as one-off cannon fodder. These group dynamics might be discerned through representations of terrorists in the inward-facing communications of the group (e.g., distinctive costumes and language) or might be coded as attributes associated with particular sources.

---

96 Domenico Tosini, "Calculated, Passionate, Pious Extremism: Beyond a Rational Choice Theory of Suicide Terrorism," Asian Journal of Social Science 38, no. 3 (2010).
97 Pratt, "Religion and Terrorism."
98 Sara M.T. Polo, "The Quality of Terrorist Violence: Explaining the Logic of Terrorist Target Choice," Journal of Peace Research 57, no. 2 (2020).
99 Daniel Koehler, "Radical Groups' Social Pressure Towards Defectors: The Case of Right-Wing Extremist Groups," Perspectives on Terrorism 9, no. 6 (2015).
100 Sageman, Understanding Terror Networks.
101 David Cook, "The Implications of "Martyrdom Operations" for Contemporary Islam," The Journal of Religious Ethics 32, no. 1 (2004).
102 Sageman, Understanding Terror Networks.

## Primary Role of Dehumanization for Distinguishing Levels 1 and 2 (Fringe Actors Versus Violent Extremists)

The violent extremist category in DMET (Level 2) includes actors (and content) that is either associated with physical violence or associated with non-physical violence in the form of dehumanization. Facebook (and by relation Instagram), Twitter, YouTube, and LinkedIn recognize dehumanization as a particularly dangerous form of hatred as it removes moral objections one may have to enact violence, even mass violence, against women,[103] children,[104] and civilians more broadly within a target group. It connects to violent extremists' cognition of representing cultural and structural violence through silencing and exclusion. It supports their group dynamics of coalescing around an out-group as the perceived or designated existential threat. While dehumanization may not always lead to violence, genocides and atrocities typically require it. This cue would identify groups or individuals that rely on dehumanizing language, or over time are spreading large amounts of dehumanizing discourse about a group identified on the basis of a protected characteristic. Dehumanization occurs in two forms:

1. Dehumanizing language includes material that presents the class of persons to have the appearance, qualities or behavior of an animal, insect, filth, form of disease or bacteria; or to be inanimate or mechanical objects; or a supernatural threat, in circumstances in which a reasonable person would conclude that the material was intended to cause others to see that class of persons as less deserving of being protected from harm or violence. This material would include words, images, and/or insignia;[105] and

2. Dehumanizing discourse or conceptions include the sustained curation of information to a specific audience to suggest that the class of persons on the basis of their identified characteristic[106]

    a. are polluting, despoiling, or debilitating society;

    b. have a diminished capacity for human warmth and feeling or independent thought;

    c. act in concert to cause mortal harm; or

    d. are to be held responsible for and deserving of collective punishment for the specific crimes, or alleged crimes of some of their "members."

---

103 Nikki Marczak, "A Century Apart: The Genocidal Enslavement of Armenian and Yazidi Women," in A Gendered Lens for Genocide Prevention, ed. Mary Michele Connellan and Christiane Fröhlich (London: Palgrave Macmillan UK, 2018).
104 Peter Lentini, "The Australian Far-Right: An International Comparison of Fringe and Conventional Politics," in The Far-Right in Contemporary Australia, ed. Mario Peucker and Debra Smith (Singapore: Springer Singapore, 2019).
105 Nick Haslam, "Dehumanization: An Integrative Review," Personality and Social Psychology Review 10, no. 3 (2006); Jonathan Leader Maynard and Susan Benesch, "Dangerous Speech and Dangerous Ideology: An Integrated Model for Monitoring and Prevention" Genocide Studies and Prevention: An International Journal 9, no. 3 (2016).
106 Haslam, "Dehumanization: An Integrative Review"; Maynard and Benesch, "Dangerous Speech and Dangerous Ideology."

While preventing dehumanization is an imperative under international law (e.g., Article 20, 2, ICCPR; Article 25, 3e of the Rome Statute) current algorithms are focused on detecting individual instances. We conceive that DMET could be trained to predict aggregate harm by specific actors from a range of samples of borderline content that each might be difficult to discern as harmful individually. Information campaigns acting as vehicles for widespread dissemination of dehumanizing conceptions and discourse will need to be distinguished from news commentary, partisan talk, or fringe discourse. We have suggested predictors to build this critical capability (discussed in 4.1).

It should be noted that the risk of violence against targeted groups is not reduced (and may be increased) when advocates are powerful voices speaking in mainstream contexts. However, where dehumanization is normative and mainstream in a regional context because it is espoused by mainstream politicians or state offices, other forms of politically- and psychologically-informed interventions or challenges may be more effective than content removal.

In our approach, such mainstream groups and content would be placed in the violent extremist category by DMET when regional norms are not considered. All the authors condemn dehumanization against any target in any context. Some authors involved in this report believe platforms could choose to downweigh such groups or content to fringe or partisan on the grounds of regional norms by using exemption functions. For example, dehumanizing homophobia, anti-Semitism, or Islamophobic dialogue advocated by mainstream actors (church leaders, politicians) might be reclassified as partisan or mainstream in certain contexts, when transparently and accountably locally normative. While the Australian Muslim Advocacy Network (AMAN) supports transparency for why certain groups or content is downweighed, it believes that downweighing of dehumanization should be avoided in any regional context by platforms to uphold the overarching obligation under international law not to contribute to the incitement of genocide.

An example of speech that would potentially trigger a violent extremist classification in the absence of regional norm adjustments is provided by political debates over introducing the death penalty for homosexuality in Uganda.[107] For example, the Ugandan Minister for Ethics and Integrity, Simon Lokodo, remarked "Homosexuality is not natural to Ugandans, but there has been a massive recruitment by gay people in schools, and especially among the youth… We want it made clear that anyone who is even involved in promotion and recruitment [of homosexuality] has to be criminalized. Those that do grave acts will be given the death sentence." Platform providers could consider regional norms despite content being flagged through DMET by transparently exempting state actors from content moderation as discussed in section 5.2 below.

---

107 "Uganda Plans to Introduce Death Penalty for Homosexuality with 'Kill the Gays' Law," ABC News, link

# Application of DMET

In the following, we illustrate the applicability of DMET by deriving potential instantiations of the cues above for the different levels of ideological engagement together with specific examples for each level. The goal is to clarify the distinction between levels while acknowledging that further efforts are necessary to identify an exhaustive set of instantiations of cues, determine cue up/downweighing methods, sharpen cut-off criteria between levels, and develop strategies to deal with issues such as the niche radicalization of splinter groups. Subsequently, we provide the integrative sample classification of organizations with various degrees of ideological engagement obtained in consultation with global experts.

## Illustrative Application of DMET

For partisanship, there would be considerable noise across contexts in how partisan contestation is expressed. At a cognitive level, our starting basket of indicators for partisanship would include simple markers of identification (e.g., use of "we," "us"), us-them distinctions (e.g., "reject," "oppose"), in-group positivity (e.g., "we are good," "we are right"), and out-group negativity (e.g., "they are wrong," "they are bad"). Particular stereotypes that are contextually relevant might be either identified via machine learning or input as cues to screen for (e.g., "Mexican gangs"). While satire and indirection create ambiguity in recognition of cues, specific contextually relevant elements could be coded (e.g., "African gangs"), alongside behavioral indicators of support for in-group actions (e.g., "donate," "volunteer") as well as politicized actions (e.g., "vote," "rally") and artistic contestation (e.g., "protest song," "protest poem"). Signals that indicate partisan group dynamics could comprise moralized grievances (e.g., "justice," "righteousness"), need for significance (e.g., "respect," "be counted") as well as victimization and crisis (e.g., "victim," "crisis," and "threat"); each could constitute initial indicators supporting categorization at this level.

Fringe groups' linguistic and image markers and beliefs would often vary contextually and require local training, but abstract indicators could include cues of dogmatism (e.g., "always," "never"), as well as moral absolutes (e.g., "hero," "villain," "traitor," and "martyr"). Behavioral indicators could include specific contextually relevant insult patterns, narratives, or more abstract categories of coding such as high-arousal negative emotions associated with the out-group-oriented, such as anger, contempt, and disgust.[108] In general, the group dynamics will not likely be transparent to content categorization, although some themes (such as purity and domination) may be available for linguistic coding. In other cases, particular sources or groups could be coded as possessing fringe-characterizing dynamics by experts, and then markers of the source group membership (e.g., jargon and group affiliation terms) could be used to identify content from the fringe

---

108 Heerdink, Koning, Doorn, and Van Kleef, "Emotions as Guardians of Group Norms."

![GIFCT - Global Internet Forum to Counter Terrorism]actors.

Violent extremists would also engage in dehumanization forms either directly in their language or through the general discourse and conceptions (as elaborated above).

In the second half of 2020, AMAN completed a study of five actors producing significant amounts of blog or pseudo-news content that triggered explicitly dehumanizing and violent responses by users on Facebook and Twitter. That study identified the following markers that were common to all five actors' information operations:

1. Dehumanizing conceptions or conspiracy theories on the actor's website (where applicable) in relation to an identified group ("the out-group") on the basis of a protected characteristic;
2. Repeated features of the headlines and images that are curated for a specific audience, including:

    - Essentializing the target identity through implicating a wide net of identities connected to the protected group (e.g., "Niqab-clad Muslima," "boat migrants," "Muslim professor," "Muslim leader," "Iran-backed jihadis," "Ilhan Omar," "Muslim father");
    - High degree of hostile verbs or actions (e.g., stabs, sets fire) attributed to those subjects;
    - Significant proportion of actor's material acting as "factual proofs" to dehumanizing conceptions about out-group;
    - Potential use of explicitly dehumanizing descriptive language (e.g., frothing-at-the-mouth) or coded extremist movement language with dehumanizing meaning (e.g., invader, a term used in RWE propaganda to refer to Muslims as a mechanically inhuman and barbaric force). However, for the most successful actors, dehumanizing slurs were avoided to maintain legitimacy and avoid detection; and
    - Where there was no dehumanizing language, there was a presence of "baiting" through rhetorical techniques like irony to provoke in-group reactions; and

3. Evidence in the user comment threads of a pattern of hate speech against the out-group.

Markers like these above could be used to train algorithms to identify an information operation intended to dehumanize an out-group over time. Further, GIFCT would be able to compile a list of protected characteristics recognized commonly by member platforms or the United Nations Strategy and Plan of Action on Hate Speech.

Violent extremists could also use images and linguistic markers of out-group violence towards the in-group and contextually relevant images and language of out-group self-

60

defense. Extremists develop narratives legitimizing violence,[109] often by framing the out-group as an enemy who is violent towards them. Hate speech and glorification of violent acts would both be indicators of this level of engagement, as we have noted above.[110] Next to reinforcing a rigid dichotomy that demands that people choose between the forces of good or evil (e.g., ISIS demands that all Sunni Muslims choose to fight with them or against them), they also use apocalyptic linguistic markers to trigger "awakening" in readers.[111] Similarly, terrorists would express beliefs about the legitimacy of killing and the glory of risking sacrificial death. Concrete incitement to violence and physically violent acts provides defining behavioral indicators of the terrorist level in DMET.

## Sample Feedback on DMET Classification of Ideologically Engaged Actors

In order to explore the applicability and feasibility of DMET, we reached out to a network of over 20 extremism researchers and counter-extremism advocates through authors' contacts. Our contacts highlighted several aspects of the framework for consideration. They highlighted the simultaneous prevalence of cues from multiple levels and the diffuse nature of some entities as movements rather than groups (e.g., Evangelical Christians). Co-occurrences were most prominent between Level 2 (Violent Extremism) and Level 3 (Terrorism) (e.g., Proud Boys, KKK). The discussion of divergent attributes and diffuse movements particularly appeared for QAnon (categorized by experts across Level 0 – 2) and Incels (Level 1 – 2). Regional differences within movements and the large in-group variability of actors, such as objecting to violence or actively engaging in violence, make movements like QAnon difficult to classify unambiguously.

Some contacts therefore pointed towards the importance of greater flexibility in the analysis. For example, they expressed that some groups would possess attributes of multiple categories (e.g., Institute of Public Affairs as borderline fringe, National Socialist Movement as borderline violent extremist). They also emphasized the multi-faceted approach of various groups assigning them to multiple types of ideological engagement (e.g., right-wing and religious: United Patriots Front; left-wing and separatists: Kurdish movements). Hence, we assume that cross-type patterns need to be acknowledged.

---

109 Pratt, "Religion and Terrorism"; Webber and Kruglanski, "The Social Psychological Makings of a Terrorist."
110 Olteanu et al., "The Effect of Extremist Violence on Hateful Speech Online"; Pratt, "Religion and Terrorism."
111 Matteo Vergani and Ana-Maria Bliuc, "The Evolution of the Isis' Language: A Quantitative Analysis of the Language of the First Year of Dabiq Magazine," SICUREZZA, TERRORISMO E SOCIETÀ 7 (2015).

| Levels of Ideological Engagement | Types of Ideological Engagement | | | | |
|---|---|---|---|---|---|
| | **Right-Wing** | **Left-Wing** | **Religious** | **Separatist** | **Single-Issue** |
| Terrorism | | | | | |
| **Level 3** Terrorism | Boogaloo Bois<br>Ku Klux Klan<br>National Socialist Network<br>The Base | Sendero Luminoso<br>Fuerzas Armadas<br>Revolucionarias de Colombia | Al-Qaeda<br>Islamic State (Daesh)<br>Jamaah Ansharut Daulah<br>Mujahidin Indonesia Timur (East Indonesia Mujahideen) | Euskadi Ta Askatasuna<br>Irish Republican Army (Provisional Irish Republican Army, Ulster Volunteer Force, Ulster Defence Association) | Army of God<br>Earth Liberation Front |
| Violent Extremism | | | | | |
| **Level 2** Violent Extremism | Blood & Honour<br>Combat 18<br>United Patriots Front (True blue crew, Lads Society)<br>Oath Keepers<br>Proud Boys<br>Jihad Watch | Antifa<br>Ejército de Liberación Nacional<br>Kurdistan Workers' Party | Forum Pembela Islam | Órganos de Resistencia Territorial | Animal Liberation Front<br>Bundy Family |
| Non-Violent Extremism | | | | | |
| **Level 1** Fringe Group | Australia First party<br>Bharatiya Janata Party<br>National Socialist Movement^ | Kurdistan Communities Union<br>Democratic Socialists of America | Brigade Manguni<br>Peoples Temple of the Disciples of Christ<br>Westboro Baptist Church | Greater Idaho Movement<br>Texas Nationalist Movement | American Family Association<br>Andha Chile |
| Non-Extremism | | | | | |
| **Level 0** Partisanship | One Nation<br>Partai Keadilan Sejahtera<br>Tea Party Movement<br>Traditionalist Worker Party | Peoples' Democratic Party (Turkey)<br>Sinn Féin | Wahdah Islamiyah | Scottish National Party | Anti-Vaxxers<br>National Abortion Rights Action League<br>No más AFP (No + AFP) |

Notes. ^ trending towards increased ideological engagement

Table 2. Tentative DMET Classification of Ideologically Engaged Actors

## Narrative Case-Based Review of Level Differences

In order to offer a hands-on illustration of the individual levels and their mutual differences, we offer a narrative review of individual cases of groups in reference to DMET.

### Level 3 (Terrorists): The Base

The DMET-based classification would, for example, echo Canada's recent decision that declared The Base as a terrorist organization. It was founded in 2018 as a neo-Nazi, white-supremacist network that describes itself as an "international survivalist and self-defense network" that seeks to train its members for fighting a race war (Counterextremism, 2021). Cognitive cues as disseminated by the group's leadership denounce the public and system as the enemy. For example, a leading member, Rinaldo Nazzaro, distributed the following message on Telegram, April 8, 2021:

> "Republicans and many White Nationalists think they're fighting for the future of America but they've already lost it and there's no hope of taking it back. The System is irreversibly dominated by the enemy…The System *is* the enemy and the enemy *is* the System—They're inherently and inseparably one and the same now. Some do realize this and hope for a spontaneous collapse which unfortunately will never come. The only victory left to be had is breaking away before it's too late."[112]

Regarding behavioral cues, The Base's leadership has called for members to focus on non-attributable actions that destabilize society. The Base has distributed to its members' manuals for lone-wolf terror attacks, bomb-making, counter-surveillance, and guerilla warfare.[113] Similarly, The Base also promotes a group dynamic with dedicated roles to engage in violent actions. For example, one post by Nazzaro on Telegram on December 21, 2020, reads

"By no later than the 90 day-mark, plan to go on the offensive by clearing and holding the nearest town. You will commandeer the town and this will serve as your new base of operations," before telling followers there may come a time where they will need to kill American citizens if their insurgency is challenged.[114]

### Level 2 (Violent Extremists): Oath Keepers

The Oath Keepers would fulfill DMET's characteristics of a violent extremist organization. They are a loosely organized collection of anti-government extremists who are part of the broader anti-government "Patriot" movement with a particular focus on recruiting current and former military members, police officers, and firefighters.[115] The Oath Keepers are driven by conspiracy theories and establish a cognitive glue that promotes violence towards the out-group by asking all members to take a pledge to oppose an allegedly tyrannical American government that will use state forces to control U.S. citizens. The pledge is targeted to refuse or disobey governmental orders to, for example, disarm the society or impose martial law.[116] Prominent Oath Keeper members such as the founder Stewart Rhodes engage in dehumanizing behavior when, for example, declaring migrants or families of legal asylum seekers as an "invasion."[117] Members of the Oath Keepers show considerable personal agency in support of their group and seek to protect its members against the out-group threat, for example by providing armed patrols during the protests in Ferguson or as armed security during land disputes. Similarly, they coalesce with the Constitutional Sheriff and Peace Officers Association (CSPOA) that disputes the federal government's authority and promotes the notion that local sheriffs do not have to obey federal authorities.[118]

---

112 "The Base," Counterextremism, Counterextremism.com, link
113 "The Base," Counterextremism.
114 "The Base," Counterextremism.
115 "The Oath Keepers," Anti-Defamation League, link

### Level 1 (Fringe Group): Westboro Baptist Church

DMET would classify the Westboro Baptist Church as a Level 1 Fringe group. The Westboro Baptist Church is an American hyper-Calvinist hate group known for engaging in inflammatory homophobic and anti-American pickets and hate speech against atheists, Jews, Muslims, transgender people, and numerous Christian denominations. The Westboro Baptist Church has an extensive indoctrination system, as evidenced by the comments of a 7-year old member towards an ABC News reporter, saying that those who were destined for eternal damnation included "gays, fags, hundreds and hundreds of Jews."[119] The group has an extensive history of engaging in denigrating antisemitic and anti-gay activities such as over 20,000 respective protests promoting the message that "Any church that allows fags to be members in good standing is a fag church […] they have created an atmosphere in this world where people believe the lie that God loves everybody."[120] The group holds and enforces strong norms of purity in their beliefs, as most pointedly described by the fact that its founder Fred Waldron Phelps Sr. was excommunicated arguably for diverging from the group's hateful demeanor by suggesting they pursue a kinder approach.[121]

### Level 0 (Partisanship): One Nation Group

DMET's transition between Level 0 and 1 appears to be more fluid than between other levels. An example of a non-extremist partisan group is the One Nation party (Pauline Hanson's One Nation). It is Australia's far-right political party that was founded in 1997. The founder Pauline Hanson promotes polarizing views of the radical right by using "us-versus-them" language. She holds political grievances that she calls "reverse-racism" or "anti-white" racism and propagates the idea of immigrants and refugees as existential threats to the safety, security, and "culture" of a particular society.[122] Behaviorally, she

---

116 "The Oath Keepers," Anti-Defamation League.
117 John Dougherty, "Oath Keepers 'Call to Action' for Flynn Sentencing a Bust," Southern Poverty Law Center, link.
118 "The Oath Keepers," Anti-Defamation League.
119 Glenn Ruppel, Kelsey Myers, and Eamon McNiff, "Raised to Hate: Kids of Westboro Baptist Church," ABC News, link
120 "Westboro Baptist Church," Anti-Defamation League, link
121 Victoria Cavaliere, "Founder of Westboro Church in Kansas Excommunicated, on Death Bed - Son," Reuters, link

64

expresses a strong populist ideology that non-natives must either assimilate and embrace "Australian culture and values" or "go back to where they came from."[123] While PHON has been described as exhibiting hate speech, calls for exclusion, and discrimination, their party obtained 10.27% of the Senate vote in Queensland in the 2019 Federal Election, double its performance nationwide. Its more mainstream acceptance or smaller deviation from the norm in Queensland may warrant its location within the level of partisanship. However, this organization would be considered fringe according to its national levels of political support (Level 1). Australia's ABC News reported that a former PHON candidate later attempted to join The Base out of frustration with the democratic system.[124]

## Illustrative Case-Based Empirical Analysis of Level Differences

Beyond the narrative review of individual cases, DMET can potentially be implemented to operationalize and systematically assess cognitive, behavioral, and group dynamic cues of movements, groups, individual actors, or content (e.g., on social media). We envision, for example, assessing social media content regarding their cognitive, behavioral, and group dynamic cues, which can then be aggregated for a particular actor (e.g., individual, group, movement). These empirical analyses can be used to quantify the profile patterns across the different levels of ideological engagements for the following purposes:

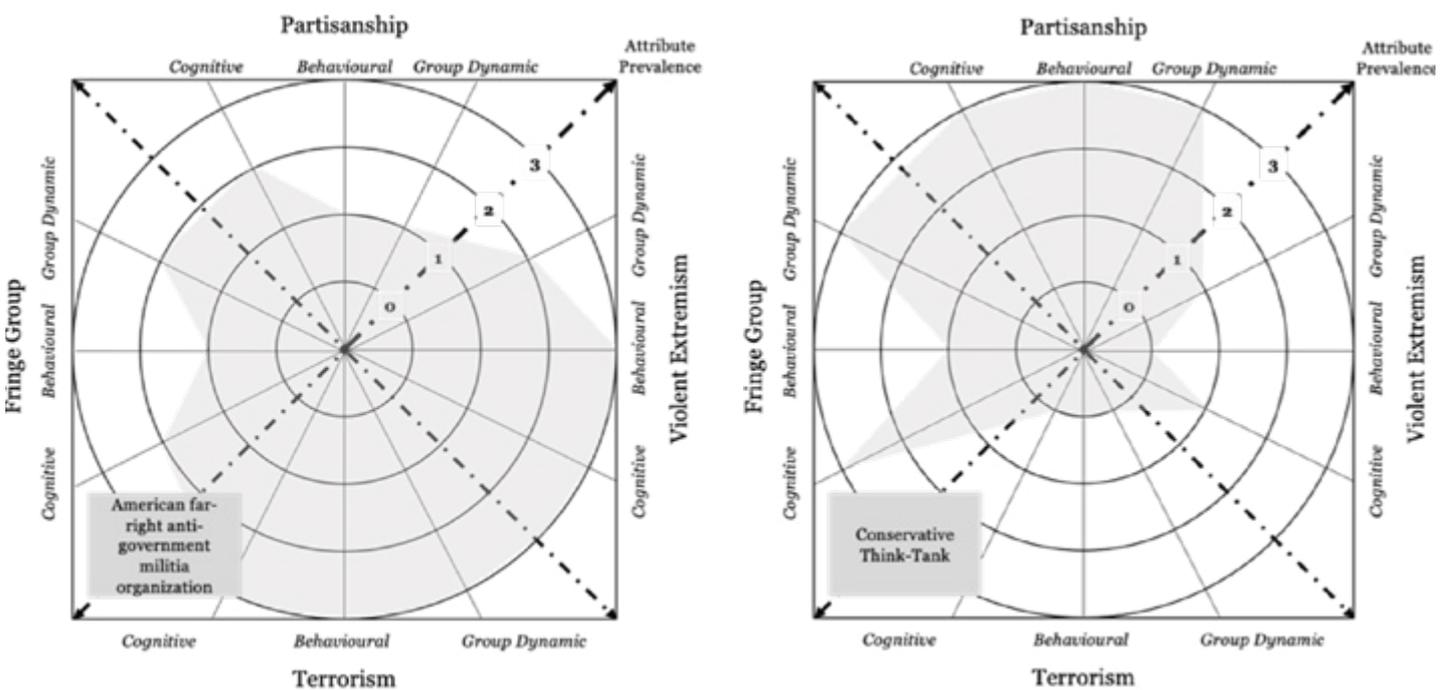

Figures 3 and 4. Conceptual actor profile-specific analyses of the prevalence of definitory cues.

---

122 Kurt Sengul, "Pauline Hanson Built a Political Career on White Victimhood and Brought Far-Right Rhetoric to the Mainstream," The Conversation, June 22, 2020, link
123 "Transcript: Pauline Hanson's 2016 Maiden Speech to the Senate," ABC News, link
124 Alex Mann and Kevin Nguyen, "The Base Tapes," ABC News, link

**Level of engagement estimation:** Given the probabilistic nature of the level defining cues, in-depth DMET-based analyses could assess the prevalence of the proposed attributes per level for different actors (Figures 3 & 4). This would help identify the potential or occurrence of splinter groups. These graphs would highlight whether an organization either solely engages, for example, in fringe activities, or whether others (under the same group name) are already engaged in terrorist activity. Decision makers can transparently assess the profile of different groups or use it to determine the potential threat (parts of) a particular group pose.

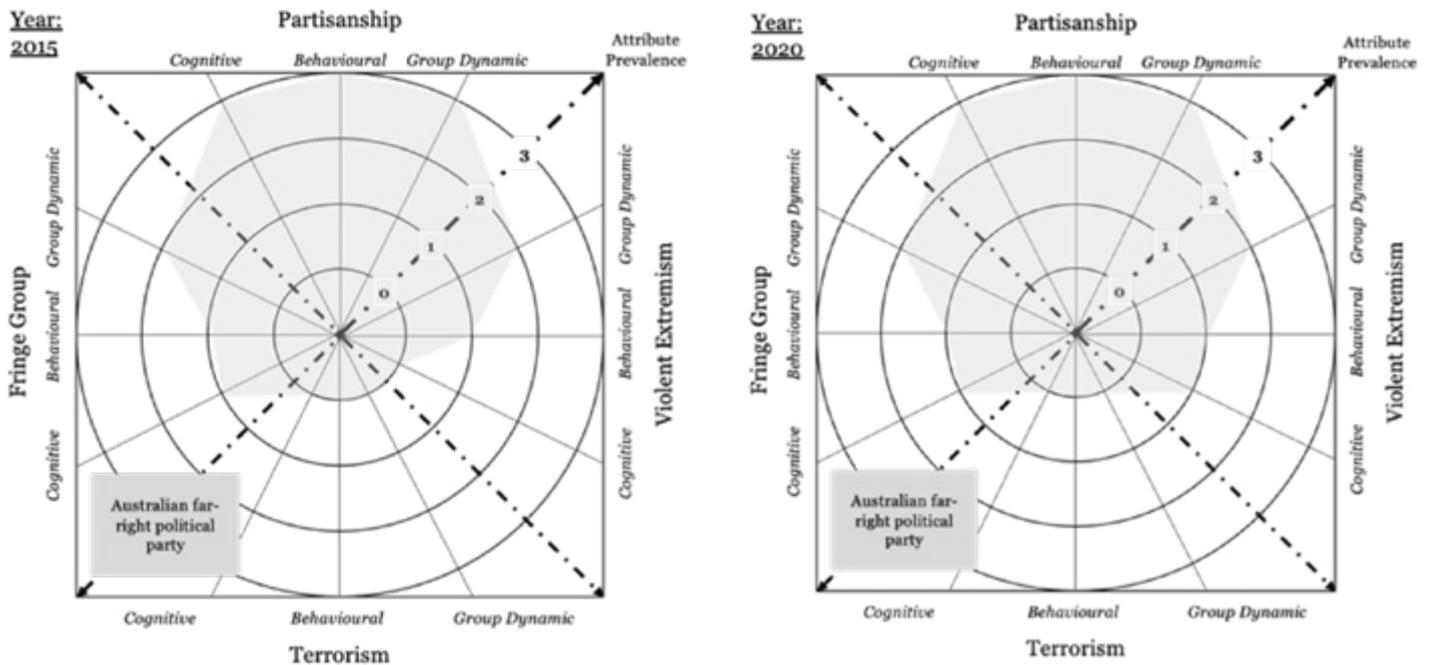

**Figures 5 and 6.** Conceptual actor profile tracking of definitory cues over time.

**Time-dependent tracking:** DMET can also be used to track and visualize changes over time (Figures 5 & 6). By operationalizing lower levels of non-violent ideological engagement (i.e., Level 0 or 1), DMET enables monitoring relevant ideologically engaged actors before they turn violent (i.e., Level 2 or 3). This could help to transparently flag suspicious actors to decision makers, issue warnings towards respective actors, and systematically re-evaluate the actor profile for de-platforming or delisting decisions.

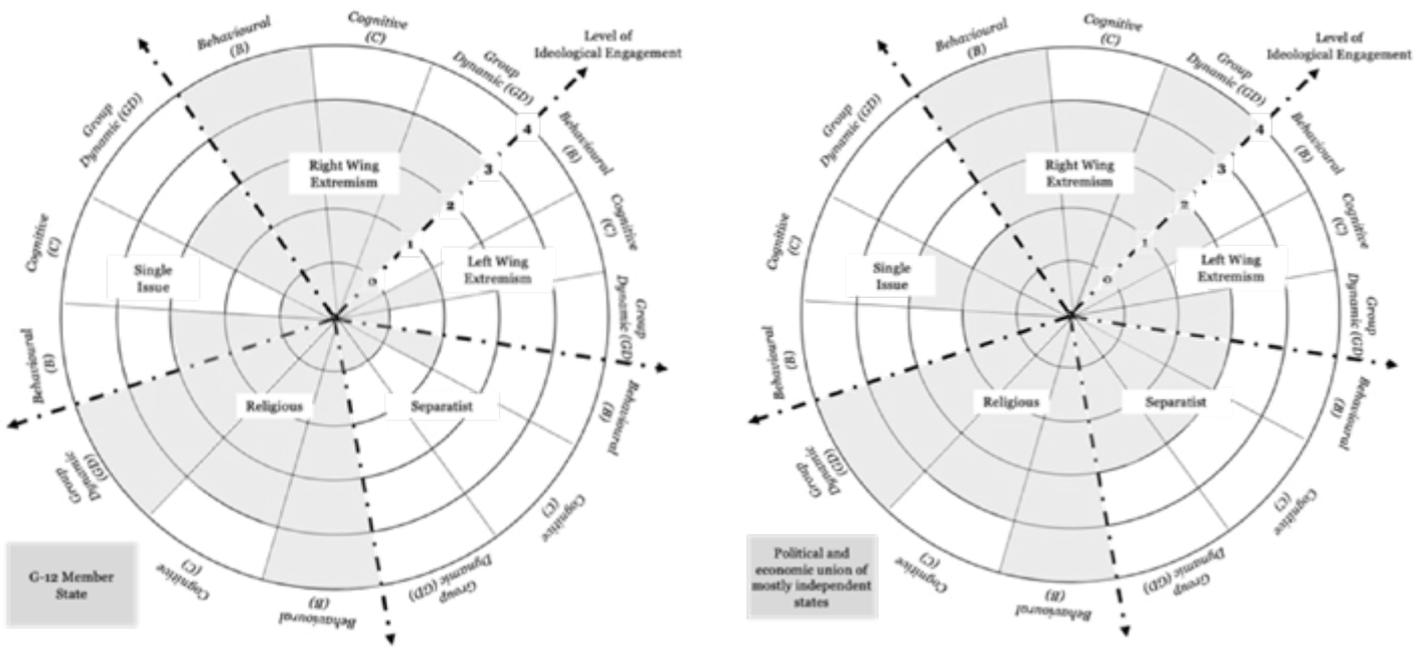

**Figures 6 and 7.** Conceptual holistic profile analyses of definitory cues across ideological types.

**Holistic profile assessment:** Furthermore, DMET can be used to assess all DMET measures for a particular entity holistically. When applied to a particular actor (i.e., individual, group, movement) this can illustrate engagement across different ideological types (e.g., religious fundamentalistic right-wing actors). It can also be applied to create a more general overview of the ideological engagement in the sense of a political barometer in a particular region (Figure 6) and across regions (e.g., state unions, Figure 7).

# Discussion

## Strategic Application Considerations of DMET

We propose DMET to be used as a foundational framework for classifying forms of extremism and associated extremist content with value beyond approval/removal decisions. We perceive DMET's vital strength in this regard to be in its adaptability to suit varying, fluctuating, or transforming perspectives on what constitutes extremisms. The value-free modularity of DMET remains unbound from fluctuating political, cultural, and/or legal perspectives and provides a depolarized snapshot of attributes on a continuum of extremist magnitudes transparent to platforms, scholars, and the broader public.

To achieve this broad perspective on the existing magnitudes of extremism, considering developments over time and geography requires comprehensive data access. This includes existing historical datasets and consultation with diverse experts and practitioners from academia and industry. This will likely be a modular, iterative approach that assures the implementation feasibility of DMET throughout. For instance, historic and geographic dimensions of extremism might not be included until later stages of the DMET development without impairment of DMET's current operability.

The modularity of DMET that accounts for different understandings and magnitudes of extremism is advantageous for its applicability into content moderation use cases. Capturing different magnitudes of extremism in a multidimensional matrix aids transparency and acts as a decision support system, especially where violent extremists share non-extremist content or vice versa. The source of the non-violent content is identified in relation to the sharing entity (i.e., a particular content's source being identified as VE Actor), aiding in making informed decisions on removing or approving content that would go unnoticed without considering these multiple dimensions of content.

The DMET proposal is qualified by the need for transparency, both prospective (transparency of design) and retrospective (transparency through inspection and explanation), and accountability (managerial and external) to secure public trust.[125] Safeguards should ensure data used to train and develop algorithms is high quality, open to academic scrutiny (including from an AI racial discrimination perspective), and continuously reviewed, corrected, and improved. Normal safeguards against algorithm discrimination in predictive policing (i.e. eliminating variables such as race and religion, or proxies for these variables) are not necessarily going to be appropriate in this context. Although some of the cues relate to cognitions, behaviors, and group dynamics that apply across the ideological spectrum, the data used to train DMET will be drawn from a range of contexts and include racial and religious terms. Checking how representative the categorization of extremists is in relation to other indicators of prevalence may provide an indicator of bias or generalizability to guide revision.

---

125 Heike Felzmann et al., "Transparency You Can Trust: Transparency Requirements for Artificial Intelligence between Legal Norms and Contextual Concerns," *Big Data & Society* 6, no. 1 (2019).

The DMET should aim to respond to published good practice,[126] principles on artificial intelligence under the Organization for Economic Co-operation and Development (OECD) that were adopted on May 22, 2019, and obligations set out in the European Commission's proposed legislation (e.g., Proposal for a Regulation of the European Parliament and of the council laying down harmonized rules on artificial intelligence, April 21, 2021), as artificial intelligence models to detect terroristic and violent extremist content will be regarded as "high-risk." Mistaken classifications promoted by GIFCT can affect users' rights to free expression on several platforms at once and can even stifle efforts to highlight human rights abuse.[127] Transparent and fair review processes, facilitated by GIFCT to quickly respond to unintended consequences, are also important. It would make it easier for human rights organizations to complain to multiple platforms at once.

We also acknowledge that violent extremist and terrorist labels are highly political and can require platforms to make exemptions for particular actors or content being flagged through DMET. In this approach, as highlighted above, platform providers can toggle content associated with particular groups (e.g., those flagged on U.N. lists) to be always categorized as terrorists, whereas other content (e.g., advocacy for violence by state actors) is not. Similarly, the collective right to self-determination is enshrined in human rights law (see Article 1.1 ICCPR as well as Article 1.1 of ICESCR), and some self-determination movements use violence. If that violence is deemed to be used in "armed conflict" by the International Committee of the Red Cross (ICRC), then humanitarian laws of war apply, and this is not treated as extremism or terrorism. However, some non-state actors will not be listed by the ICRC and governed by humanitarian law because they are engaged in a conflict that does not meet threshold tests.[128] For example, some violent protests will be flagged in DMET, including protests where protestors use violence in response to (or to resist) state violence, including physical violence and life-threatening structural violence. To contend with this gap, platforms could continue to have the discretion to exempt further non-state actors based on exercising self-defense,[129] considering principles such as self-determination, duress, necessity, proportionality, or on the balance of other fundamental human rights. Our hope would be that if a group's content is flagged in DMET for the use of dehumanization or advocacy of violence but exempted by platforms, they would be transparent about exemptions made and provide reasons.

Making a decision based on the balance of fundamental human rights may be required to provide an enduring mechanism for managing conflict between differing world views and claims. For example, groups that wish to express themselves, their beliefs, and exercise their fundamental rights (such as the right to parent their children and choose schooling according to their beliefs) are protected by human rights law to do so. Groups are not protected to infringe upon the fundamental rights of others, and such behavior would begin to animate DMET cues.

---

126 Felzmann et al., "Transparency You Can Trust."
127 Abdul Rahman Al Jaloud et al., "Caught the Net: The Impact of "Extremist" Speech Regulations on Human Rights Content," Electronic Frontier Foundation ed. JIllian C. York (2019), link
128 See the categorization of armed conflict as proposed by UNODC: link
129 Ben Saul, "Defending 'Terrorism': Justifications and Excuses for Terrorism in International Criminal Law," Australian Yearbook of International Law 25 (2008).

Noting that DMET focuses on deviation from social norms, it is important to consider that a violent protest will deviate less from mainstream social norms in some regional contexts. For example, where mass popular protest movements feature violent elements and advocacy of violence against law enforcement and the state, the scale of people involved will mean that their behavior may not be flagged as non-normative, extreme, or radical by regional standards.

## Technical Implementation Considerations of DMET

In our approach, large baskets of indicators would be associated probabilistically with each level (e.g., cognitive stereotypes, dehumanizing language, calls for violence) and will be used to develop models that algorithmically classify groups (or content) into categories, with each cue weighed according to its ability to discriminate in particular contexts defined by the other cues as well as input from the platform provider where desired. We imagine an incoming stream of content coded for the indicator cues and the groups involved via machine learning in a process that would be more error-prone at first and be refined over time and regionally to produce context-specific accuracy, which would in turn decay as diagnostic attributes changed over time until relearned dynamically. Especially in the beginning, this will require extensive human oversight, for example in order to deal with expected inaccuracies of automated machine learning algorithms when dealing with linguistic markers for irony, sarcasm, or subtle dehumanization.

Content and groups would be classified probabilistically into categories where cues have established sufficient discriminant validity and high confidence (e.g., with explicit calls to violence, or when the content is sourced from a group identified as a terrorist organization, or as a state actor or journalistic or academic source). More commonly, groups and individuals would be classified based on profiles established via multiple content posts with increasing confidence over time, with each content item or group reciprocally associated with transparent certainty/uncertainty scores based on a profile of attributes, which could be available to platforms as an output.

As a next step, DMET could train a machine learning algorithm that identifies the different level cues in online content. Due to the linguistic challenges and subjective interpretation of the content in this unique context, unsupervised learning approaches are likely to provide misleading results. There is a need for more sophisticated models, and building such models requires generating a labeled training data set from scratch. A potential cold start problem of insufficient data for initially training the algorithm could be overcome by collecting social media content from the groups, building on those identified by experts as above. This data could comprise social media posts and comments, memes, or content from external sites (e.g., extremist websites, blogs) introduced into conversations through hyperlinks. The accuracy of the machine learning algorithm will mainly depend on the (1) clarity and consistency of classification rule (the coding scheme), (2) quality and size of labeled data, and (3) finding the proper feature representation.

Subsequently, we could analyze the actor-specific distribution of individual messages across the different levels to establish and identify communication patterns. By applying the previously developed machine learning algorithm to new data, we could expand the available content coded data. This would help perform a ROC analysis to determine extremism cut-off scores between levels. The regression weights for the individual cues could also serve as an indicator for the up-/downweighing of individual cues. By choosing to downweigh or upweigh particular dimensions, platform providers can establish local profiles of tolerance (e.g., no hate speech at all versus this group; versus hate speech tolerated against this group, due to its being normative in this context) in a way that is transparent and able to be accountable or engaged with dialogue. Platform providers may also opt for transparently and accountably in exempting certain actors such as state organizations or religious groups ex officio because their views (e.g., the Ugandan minister above) are mainstream rather than extreme in the regional context.

We could complement the analysis by using metadata that contains information on the connection between entities to consider the hierarchical structure between individual extremists (potentially) embedded in extremist groups who are themselves nested in ideological movements.

## Benefits of DMET

We understand extremism as a dimensional concept, with terrorism as the most deviant pole from the regional norm. DMET supports:

- **Ideological fairness:** equal opportunity for all entities to be classified as terrorist based on the generally applicable cognitive, behavioral, and group dynamic cues;
- **Global applicability and scalability:** definitory cues transcend geographical, cultural, and political borders and can be applied relative to relevant reference norms;
- **Update frequency:** Observable changes in cognitions, behaviors, or group dynamics can be captured through near real-time updates;
- **Transparency:** Classifying actors according to DMET categories based on their degree of ideological engagement enables transparency and accountability in regulatory decisions;
- **Surfacing states' role and influence:** Current definitions of extremists or terrorists

often exclude state actors. DMET potentially classifies any kind of actor without consideration of their societal role. Platforms can navigate these situations by making transparent exemptions in reference to the classifications proposed by DMET to justify their decision making;

- **Reduced probability of misclassification errors for (non-) violent extremisms:** A more nuanced understanding of the degrees of ideological engagement and the potential sub-groups reduces the probability of wrong decisions (classifying regular users/content as terrorist or failing to identify terrorists as such); and

- **Attention to violence in all its forms:** Many existing legal frameworks are so piecemeal or narrow that they deprioritize and overlook the experience of victims and communities targeted by terroristic and violent extremist violence. DMET contemplates the full continuum of violence that occurs in the violent denial of diversity, including structural and psychological violence. Importantly, it recognizes serial or systematic dehumanization of an out-group as an attribute of violent extremism.

Subsequently, DMET addresses contemporary challenges of extremism classification and associated content moderation approaches, including the lack of consent on universal definitions of extremism, bias, and deficient objectivity on different magnitudes of extremism.[130] DMET's multidimensional approach enables the aggregation of various lists and dimensions to allow biases in views of extremism (e.g., exemptions for certain actors) to be more transparent and accountable. Moreover, DMET unlocks the possibility of a 360-degree context view of extremism irrespective of the limitations of individual extremism lists and allows for tracking the development in terms of (de-)radicalization over time through continuous assessments among the spectrum of ideological engagement.

## Boundary Conditions of DMET

- Dimensionality of attributes: DMET ascribes attributes to particular levels to capture different degrees of severity. It needs to be acknowledged that these attributes themselves can also be dimensional (e.g., expressing blame for negative events can be more or less rampant). Similarly, actors might, for example, dehumanize a group by using a multitude of cues that holistically dehumanize the target without making it explicit in one singular instance. The dimensionality of attributes needs to be empirically assessed to be statistically considered through measures of item difficulty and item discrimination. The individually classified instances then need to be holistically considered for each entity. This would also enable DMET classifications to record an actor's tendency of either trending towards a higher (or lower) DMET level of ideological

---

130   Meserole and Byman, "Terrorist Definitions and Designations Lists."

engagement or of being stable within the level.

- Probability of attributes: Similarly, DMET ascribes attributes to levels of ideological engagement where we expect their highest probability of prevalence. We acknowledge that ideologically engaged actors (i.e., individuals, groups, movements) can simultaneously demonstrate cognitive, behavioral, and group dynamic cues from multiple levels or not express particular cues from the level at which they are classified. The classification of ideological actors according to DMET requires an empirical assessment of the expectable probabilities of cues per level and their respective level-determinant weight (i.e., up- or downweighing of attributes).

- Combinability of types: While DMET distinguishes five common types of ideological engagement, we understand that actors (i.e., individuals, groups, movements) can simultaneously follow different types of ideologies (e.g., nationalism in combination with religious fundamentalism). Hence, DMET classifications need to acknowledge the expressivity of characteristics across multiple ideological types per actor.

- Fragmentation of actors: Different actor organizations (i.e., groups, movements) can include splinter groups or individuals that diverge from the characteristics of the overall organization (e.g., enact or support violent behavior as opposed to the general movement). These individuals or groups can either emerge as splinter groups or lone-wolf actors alongside the general movement or relative to their location (e.g., violent in one country, non-violent in another). Classifications according to DMET need to acknowledge the relatedness of the individual groups or actors to the higher-level organization (e.g., via metadata) while considering their individual particularities.

- Time and cultural specificity: Actors (i.e., individuals, groups, movements) as well as the expression of cues evolve. Actors might become more or less ideologically engaged, splinter-off into different sub-groups, or form coalitions with other movements. Similarly, how cues are expressed can change as words can adopt different meanings over time, as the societal acceptability of terms evolves, or as terms have regionally specific meanings (e.g., reference to Odin in Nordic nationalist groups versus the rest of the world). DMET classifications need to be considered at particular points in time, in particular regions, and regularly re-evaluated in predetermined time intervals (e.g., to inform de-platforming or readmittance and delisting decisions).

- Complexity of societal norms: DMET characteristics need to be assessed against relevant societal norms, which is intended to support its general applicability. However, regional norms may vary substantially.

# Conclusion

This paper has put forward a proposal for extending the binary understanding of terrorism (versus non-terrorism) with a Dynamic Matrix of Extremisms and Terrorism (DMET) that builds upon the notion of an underlying continuum of ideological engagement to address the intersection of extremism types (and associated extremist content) that lead to ineffective content tagging. DMET considers different types of ideological engagement and different levels, identified using cognitive and behavioral attributes and attributes of group dynamics. DMET is dynamic as it can be adapted to accommodate region- and time-specific notions of ideological engagement. The goal of DMET is to enable platform providers to make transparent and accountable decisions about engaging with content and groups so that violent extremist and terrorist content can be identified in a way that makes explicit the criteria and dimensions underlying the categorization and allows areas of contestation and change to be identified.

DMET Graphics and Visualizations

# Reference

ABC News. "Transcript: Pauline Hanson's 2016 Maiden Speech to the Senate." ABC News, [link](link)

———. "Uganda Plans to Introduce Death Penalty for Homosexuality with 'Kill the Gays' Law." ABC News, [link](link)

Abrams, Dominic, José M Marques, Nicola Bown, and Michelle Henson. "Pro-Norm and Anti-Norm Deviance within and between Groups." Journal of personality and social psychology 78, no. 5 (2000): 906.

Abrams, Dominic, Giovanni A Travaglino, José M Marques, Ben Davies, and Georgina Randsley de Moura. "Collective Deviance: Scaling up Subjective Group Dynamics to Superordinate Categories Reveals a Deviant Ingroup Protection Effect." Journal of Personality and Social Psychology (2021).

Ackerman, Gary A, Jun Zhuang, and Sitara Weerasuriya. "Cross-Milieu Terrorist Collaboration: Using Game Theory to Assess the Risk of a Novel Threat." Risk analysis 37, no. 2 (2017): 342-71.

ADL. "The Oath Keepers " Anti Defamation League, [link](link)

———. "Westboro Baptist Church." Anti-Defamation League, [link](link)

Aytaç, S Erdem, Ali Çarkoğlu, and Ezgi Elçi. "Partisanship, Elite Messages, and Support for Populism in Power." European Political Science Review 13, no. 1 (2021): 23-39.

Bakker, Bert N., Yphtach Lelkes, and Ariel Malka. "Understanding Partisan Cue Receptivity: Tests of Predictions from the Bounded Rationality and Expressive Utility Perspectives." The Journal of Politics 82, no. 3 (2020): 1061-77.

Bandura, Albert. "Moral Disengagement in the Perpetration of Inhumanities." Personality and Social Psychology Review 3, no. 3 (1999/08/01 1999): 193-209.

Baumeister, Roy F. Evil: Inside Human Cruelty and Violence. WH Freeman/Times Books/Henry Holt & Co, 1996.

Berger, John M. "Deconstruction of Identity Concepts in Islamic State Propaganda: A Linkage-Based Approach to Counter-Terrorism Strategic Communications." 1-21. The Hague, Netherlands: EUROPOL, 2017.

Borum, Randy. "Radicalization into Violent Extremism I: A Review of Social Science Theories." Journal of strategic security 4, no. 4 (2011): 7-36.

Cavaliere, Victoria. "Founder of Westboro Church in Kansas Excommunicated, on Death Bed - Son." Reuters, [link](link)

Christmann, Kris. "Preventing Religious Radicalisation and Violent Extremism: A Systematic Review of the Research Evidence." (2012).

Cook, David. "The Implications of "Martyrdom Operations" for Contemporary Islam." The Journal of Religious Ethics 32, no. 1 (2004): 129-51.

Coulson, Andrew. "Education and Indoctrination in the Muslim World." Policy Analysis 29 (2004): 1-36.

Counterextremism. "The Base." Counterextremism.com, [link](link)

Dougherty, John. "Oath Keepers 'Call to Action' for Flynn Sentencing a Bust." Southern Povery Law Center, [link](link)

Feddes, Allard R., Lars Nickolson, Liesbeth Mann, and Bertjan Doosje. Psychological Perspectives of Radicalization. London: Routledge, 2020.

Felzmann, Heike, Eduard Fosch Villaronga, Christoph Lutz, and Aurelia Tamò-Larrieux. "Transparency You Can Trust: Transparency Requirements for Artificial Intelligence between Legal Norms and Contextual Concerns." Big Data & Society 6, no. 1 (2019): 2053951719860542.

Giner-Sorolla, Roger, Bernhard Leidner, and Emanuele Castano. "Dehumanization, Demonization, and Morality Shifting: Paths to Moral Certainty in Extremist Violence." Extremism and the psychology of uncertainty (2012): 165-82.

Goffman, Erving. Stigma: Notes on the Management of Spoiled Identity. Simon and Schuster, 2009.

Graham, Roderick. "Inter-Ideological Mingling: White Extremist Ideology Entering the Mainstream on Twitter." Sociological Spectrum 36, no. 1 (2016/01/02 2016): 24-36.

Haslam, Nick. "Dehumanization: An Integrative Review." Personality and Social Psychology Review 10, no. 3 (2006): 252-64.

Heerdink, Marc W, Lukas F Koning, Evert A Van Doorn, and Gerben A Van Kleef. "Emotions as Guardians of Group Norms: Expressions of Anger and Disgust Drive Inferences About Autonomy and Purity Violations." Cognition and emotion 33, no. 3 (2019): 563-78.

Hogg, M. A., and S. A. Reid. "Social Identity, Self-Categorization, and the Communication of Group Norms." [In English]. Communication Theory 16, no. 1 (Feb 2006): 7-30.

Hogg, M. A., and D. J. Terry. "Social Identity and Self-Categorization Processes in Organizational Contexts." [In English]. Academy of Management Review 25, no. 1 (Jan 2000): 121-40.

Hogg, Michael A, Arie Kruglanski, and Kees Van den Bos. "Uncertainty and the Roots of Extremism." Journal of Social Issues 69, no. 3 (2013): 407-18.

Holbrook, Donald. "Designing and Applying an 'Extremist Media Index'." Perspectives on Terrorism 9, no. 5 (2015): 57-68.

———. "Far Right and Islamist Extremist Discourses: Shifting Patterns of Enmity." Extreme Right Wing Political Violence and Terrorism (2013): 215-37.

Horgan, John G, Max Taylor, Mia Bloom, and Charlie Winter. "From Cubs to Lions: A Six Stage Model of Child Socialization into the Islamic State." Studies in Conflict & Terrorism 40, no. 7 (2017): 645-64.

Jaloud, Abdul Rahman Al, Hadi Al Khatib, Jeff Deutch, Dia Kayyali, and Jillian C. York. "Caught the Net: The Impact of "Extremist" Speech Regulations on Human Rights Content." edited by JIllian C. York, 2019.

Jegede, Ayodele Samuel. "What Led to the Nigerian Boycott of the Polio Vaccination Campaign?" [In eng]. PLoS medicine 4, no. 3 (2007): e73-e73.

Keatinge, Tom, and Florence Keen. "Social Media and (Counter) Terrorist Finance: A Fund-Raising and Disruption Tool." Studies in Conflict & Terrorism 42, no. 1-2 (2019/02/01 2019): 178-205.

Keatinge, Tom, Florence Keen, and Kayla Izenman. "Fundraising for Right-Wing Extremist Movements." The RU.S.I Journal 164, no. 2 (2019/02/23 2019): 10-23.

Kinder, Donald R, and D Roderick Kiewiet. "Economic Discontent and Political Behavior: The Role of Personal Grievances and Collective Economic Judgments in Congressional Voting." American Journal of Political Science (1979): 495-527.

Knobloch-Westerwick, Silvia, and Simon M Lavis. "Selecting Serious or Satirical, Supporting or Stirring News? Selective Exposure to Partisan Versus Mockery News Online Videos." Journal of Communication 67, no. 1 (2017): 54-81.

Koehler, Daniel. "Radical Groups' Social Pressure Towards Defectors: The Case of Right-Wing Extremist Groups." Perspectives on Terrorism 9, no. 6 (2015): 36-50.

Kruglanski, Arie W, Katarzyna Jasko, Marina Chernikova, Michelle Dugas, and David Webber. "To the Fringe and Back: Violent Extremism and the Psychology of Deviance." American Psychologist 72, no. 3 (2017): 217.

Kruglanski, Arie W., Michele J. Gelfand, Jocelyn J. Bélanger, Anna Sheveland, Malkanthi Hetiarachchi, and Rohan Gunaratna. "The Psychology of Radicalization and Deradicalization: How Significance Quest Impacts Violent Extremism." Political Psychology 35 (2014): 69-93.

Lentini, Peter. "The Australian Far-Right: An International Comparison of Fringe and Conventional Politics." In The Far-Right in Contemporary Australia, edited by Mario Peucker and Debra Smith, 19-51. Singapore: Springer Singapore, 2019.

Lupu, Noam. "Party Polarization and Mass Partisanship: A Comparative Perspective." Political Behavior 37, no. 2 (2015/06/01 2015): 331-56.

Mandel, David R. "The Role of Instigators in Radicalization to Violent Extremism." Psychosocial, Organizational, and Cultural Aspects of Terrorism: Final Report to NATO HFM140/RTO. Brussels: NATO (2011).

Mann, Alex, and Kevin Nguyen. "The Base Tapes." ABC News, [link](#)

Marczak, Nikki. "A Century Apart: The Genocidal Enslavement of Armenian and Yazidi Women." In A Gendered Lens for Genocide Prevention, edited by Mary Michele Connellan and Christiane Fröhlich, 133-62. London: Palgrave Macmillan UK, 2018.

Maynard, Jonathan Leader, and Susan Benesch. "Dangerous Speech and Dangerous Ideology: An Integrated Model

for Monitoring and Prevention ." Genocide Studies and Prevention: An International Journal 9, no. 3 (2016): 70-95.

McCauley, Clark, and Sophia Moskalenko. "Mechanisms of Political Radicalization: Pathways toward Terrorism." Terrorism and political violence 20, no. 3 (2008): 415-33.

Meserole, Chris, and Daniel L. Byman. "Terrorist Definitions and Designations Lists What Technology Companies Need to Know." In Global Research Network on Terrorism and Technology, 2019.

Moghaddam, F. M. "The Staircase to Terrorism a Psychological Exploration." American Psychologist 60, no. 2 (2005): 161-69.

Morrison, Sara. "Westboro Baptist Church's Founder Is Dying, Excommunicated." The Atlantic, [link](#)

Neumann, Peter R. "The Trouble with Radicalization." International affairs 89, no. 4 (2013): 873-93.

Oegema, Dirk, and Bert Klandermans. "Why Social Movement Sympathizers Don't Participate: Erosion and Nonconversion of Support." American Sociological Review (1994): 703-22.

Olteanu, Alexandra, Carlos Castillo, Jeremy Boy, and Kush Varshney. "The Effect of Extremist Violence on Hateful Speech Online." Paper presented at the Proceedings of the International AAAI Conference on Web and Social Media, 2018.

Polo, Sara M.T. "The Quality of Terrorist Violence: Explaining the Logic of Terrorist Target Choice." Journal of peace research 57, no. 2 (2020): 235-50.

Pratt, Douglas. "Religion and Terrorism: Christian Fundamentalism and Extremism." Terrorism and Political Violence 22, no. 3 (2010): 438-56.

Ruppel, Glenn, Kelsey Myers, and Eamon McNiff. "Raised to Hate: Kids of Westboro Baptist Church." ABC News, [link](#)

Sageman, Marc. Understanding Terror Networks. University of Pennsylvania press, 2011.

Saul, Ben. "Defending 'Terrorism': Justifications and Excuses for Terrorism in International Criminal Law." Australian Yearbook of International Law 25 (2008): 177-226.

Schmid, Alex P. "Violent and Non-Violent Extremism: Two Sides of the Same Coin." International Centre for Counter-Terrorism (ICCT) Research Paper (2014): 1-29.

Sengul, Kurt. "Pauline Hanson Built a Political Career on White Victimhood and Brought Far-Right Rhetoric to the Mainstream." June 22, 2020 2020.

Sidanius, J., C. Van Laar, S. Levin, and S. Sinclair. "Ethnic Enclaves and the Dynamics of Social Identity on the College Campus: The Good, the Bad, and the Ugly." [In English]. Journal of Personality and Social Psychology 87, no. 1 (Jul 2004): 96-110.

Sides, John, Michael Tesler, and Lynn Vavreck. Identity Crisis: The 2016 Presidential Campaign and the Battle for the Meaning of America. Princeton University Press, 2019.

Sternberg, Robert J. "A Duplex Theory of Hate: Development and Application to Terrorism, Massacres, and Genocide." Review of General Psychology 7, no. 3 (2003): 299-328.

Stott, C., P. Hutchison, and J. Drury. "'Hooligans' Abroad? Inter-Group Dynamics, Social Identity and Participation in Collective 'Disorder' at the 1998 World Cup Finals." [In English]. British Journal of Social Psychology 40 (Sep 2001): 359-84.

Tosini, Domenico. "Calculated, Passionate, Pious Extremism: Beyond a Rational Choice Theory of Suicide Terrorism." Asian Journal of Social Science 38, no. 3 (2010): 394-415.

van Zomeren, M., T. Postmes, and R. Spears. "Toward an Integrative Social Identity Model of Collective Action: A Quantitative Research Synthesis of Three Socio-Psychological Perspectives." [In English]. Psychological Bulletin 134, no. 4 (Jul 2008): 504-35.

Vergani, Matteo, and Ana-Maria Bliuc. "The Evolution of the Isis' Language: A Quantitative Analysis of the Language of the First Year of Dabiq Magazine." SICUREZZA, TERRORISMO E SOCIETÀ 7 (01/01 2015).

Verkuyten, M., and A. De Wolf. "The Development of in-Group Favoritism: Between Social Reality and Group Identity." [In English]. Developmental Psychology 43, no. 4 (Jul 2007): 901-11.

Victoroff, Jeff, Janice R. Adelman, and Miriam Matthews. "Psychological Factors Associated with Support for Suicide Bombing in the Muslim Diaspora." Political Psychology 33, no. 6 (2012): 791-809.

Vinciarelli, Alessandro, Maja Pantic, Hervé Bourlard, and Alex Pentland. "Social Signal Processing: State-of-the-Art and Future Perspectives of an Emerging Domain." In Proceedings of the 16th ACM international conference on Multimedia, 1061–70. Vancouver, British Columbia, Canada: Association for Computing Machinery, 2008.

Webber, David, and Arie W Kruglanski. "The Social Psychological Makings of a Terrorist." Current opinion in psychology 19 (2018): 131-34.

West, Emily A, and Shanto Iyengar. "Partisanship as a Social Identity: Implications for Polarization." Political Behavior (2020): 1-32.

Wetts, Rachel, and Robb Willer. "Who Is Called by the Dog Whistle? Experimental Evidence That Racial Resentment and Political Ideology Condition Responses to Racially Encoded Messages." Socius 5 (2019): 2378023119866268.

White, Jonathan, and Lea Ypi. The Meaning of Partisanship. Oxford University Press, 2016.

Wibisono, Susilo, Winnifred R Louis, and Jolanda Jetten. "A Multidimensional Analysis of Religious Extremism." Frontiers in psychology 10 (2019): 2560.

Winter, Charlie, Peter Neumann, Alexander Meleagrou-Hitchens, Magnus Ranstorp, Lorenzo Vidino, and Johanna Fürst. "Online Extremism: Research Trends in Internet Activism, Radicalization, and Counter-Strategies." International Journal of Conflict and Violence (IJCV) 14, no. 2 (2020): 1-20.

Wintrobe, Ronald. Rational Extremism: The Political Economy of Radicalism. Cambridge University Press, 2006.

Wright, S. C., D. M. Taylor, and F. M. Moghaddam. "Responding to Membership in a Disadvantaged Group - from Acceptance to Collective Protest." [In English]. Journal of Personality and Social Psychology 58, no. 6 (Jun 1990): 994-1003.

To learn more about the Global Internet Forum to Counter Terrorism (GIFCT), please visit our website or email outreach@gifct.org

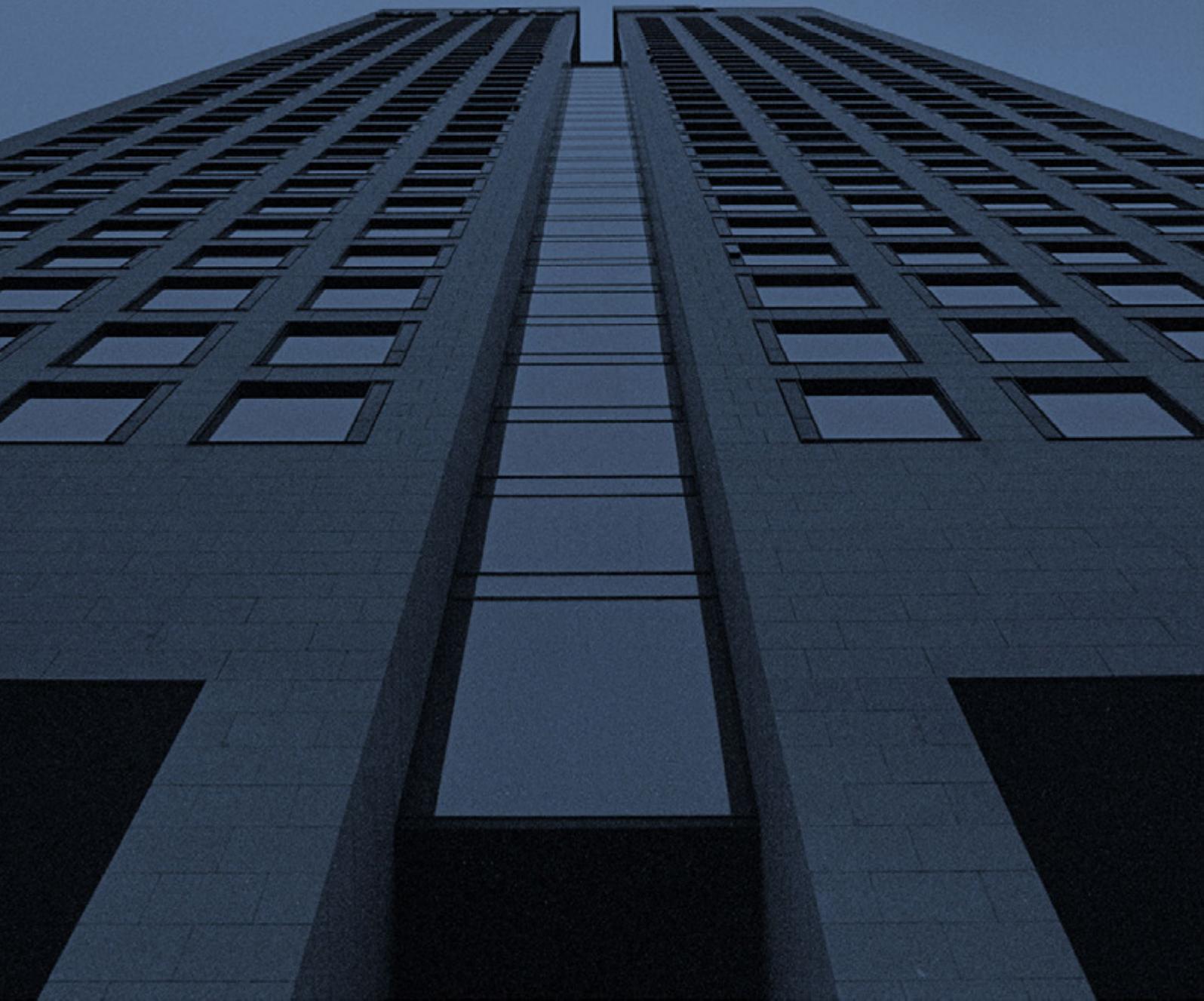